\documentclass[useAMS,usenatbib]{mn2e}

\usepackage{graphicx,natbib, ae, aecompl, amsmath, bm, amssymb, longtable}
\usepackage[T1]{fontenc}
\allowdisplaybreaks

\newcommand{\Tstart}{T_{\text{start}}}
\newcommand{\Tobs}{T_{\text{obs}}}
\newcommand{\Tlag}{T_{\text{lag}}}

\title[Cross-correlation search for continuous-wave gravitational radiation from a neutron star in SNR 1987A]{Designing a cross-correlation search for continuous-wave gravitational radiation from a neutron star in the supernova remnant SNR 1987A}
\author[C. T. Y. Chung, A. Melatos, B. Krishnan, J. T. Whelan]{C. T. Y. Chung$^{1}$, A. Melatos$^{1}$, B. Krishnan$^{2}$, J. T. Whelan$^{3}$\\
$^{1}$School of Physics, University of Melbourne, Parkville, VIC 3010, Australia\\
$^{2}$Max Planck Institut für Gravitationsphysik, Am Mühlenberg 1, D-14476 Golm, Germany\\
$^{3}$Center for Computational Relativity and Gravitation and School of Mathematical Sciences, \\
Rochester Institute of Technology, 85 Lomb Memorial Drive, Rochester, NY 14623, USA\\
LIGO-P1000089-v3 }

\begin{document}

\date{ }

\pagerange{\pageref{firstpage}--\pageref{lastpage}} \pubyear{2010}

\maketitle

\label{firstpage}

\begin{abstract}
A strategy is devised for a semi-coherent cross-correlation search for a young neutron star in the supernova remnant SNR 1987A, using science data from the Initial LIGO and/or Virgo detectors. An astrophysical model for the gravitational wave phase is introduced which describes the star's spin down in terms of its magnetic field strength $B$ and ellipticity $\epsilon$, instead of its frequency derivatives. The model accurately tracks the gravitational wave phase from a rapidly decelerating neutron star under the restrictive but computationally unavoidable assumption of constant braking index, an issue which has hindered previous searches for such young objects. The theoretical sensitivity is calculated and compared to the indirect, age-based wave strain upper limit. The age-based limit lies above the detection threshold in the frequency band 75\,Hz $\lesssim \nu \lesssim 450$\,Hz. The semi-coherent phase metric is also calculated and used to estimate the optimal search template spacing for the search. The range of search parameters that can be covered given our computational resources ($\sim 10^9$ templates) is also estimated.  For Initial LIGO sensitivity, in the frequency band between 50\,Hz and 500\,Hz, in the absence of a detected signal, we should be able to set limits of $B \gtrsim 10^{11}$\,G and $\epsilon \lesssim 10^{-4}$.
\end{abstract}


\section{Introduction}
\label{sec:ccintro}
The Laser Interferometer Gravitational Wave Observatory (LIGO) achieved its design sensitivity during its fifth science run [S5; \citep{lscinstrument09}]. Analysis of S5 data is progressing well, with new upper limits being placed on the strength of various classes of burst sources \citep{lscburst09, lscburst10, lscburst10b}, stochastic backgrounds \citep{giampanis08, lscstochastic09}, compact binary sources \citep{lsccbc09, lscinspiral10, lsccbc10} and continuous-wave sources \citep{allsky09, lsceah09, knownpul09, casa10}. In some cases, the LIGO limits on astrophysical parameters beat those inferred from electromagnetic astronomy, e.g. the maximum ellipticity and internal magnetic field strength of the Crab pulsar \citep{crab08, knownpul09}. Recently, an S5 search was completed which placed upper limits on the amplitude of $r$-mode oscillations of the neutron star in the supernova remnant Cassiopeia A \citep{casa10}. 

Aspherical, isolated neutron stars constitute one promising class of continuous-wave source candidates \citep{ostriker69}. The origin of the semi-permanent quadrupole in these objects can be thermoelastic \citep{melatos00, ushomirsky00, nayyar06, haskell08} or hydromagnetic \citep{bonazzola96, cutler02, haskellb08, haskell08, akgun08, mastrano10}. Thermoelastic deformations arise due to uneven electron capture rates in the neutron star crust. A persistent 5\% temperature gradient at the base of the crust produces a mass quadrupole moment of $\sim 10^{38}$\,g cm$^{2}$ \citep[$\epsilon \sim 10^{-7}$;][]{ushomirsky00}. Hydromagnetic deformations, on the other hand, are produced by large internal magnetic fields, and misaligned magnetic and spin axes. For example, a neutron star with spin frequency 300\,Hz and internal toroidal field $B_t \gtrsim 3.4 \times 10^{12}$\,G has an ellipticity $\epsilon \sim 10^{-6} \left(\langle B_t \rangle / 10^{15}\,\text{G}\right)$ \citep{cutler02}. The deformation of an ideal fluid star with an arbitrary magnetic field distribution and a barotropic equation of state can be computed, with ellipticities as high as $10^{-5}$ predicted for some configurations \citep{haskellb08}. Additionally, some exotic neutron star models (e.g. solid strange quark stars) allow for ellipticities as large as $10^{-4}$ \citep{owen05}. Accreting neutron stars in binary systems can form magnetic mountains, with $\epsilon \leq 10^{-5}$ \citep{melpa05, vigmel09}.
Deformations of all kinds relax viscoelastically and resistively over time, so that  young neutron stars are expected to be generally stronger gravitational wave emitters. For example, thermoelastic deformations relax on the thermal conduction time-scale ($\sim 10^4$\,yr), after the temperature gradient in the crust has switched off \citep{brown98, vigmel10}. Magnetic mountains relax as the accreted matter diffuses through the magnetic field on the Ohmic time-scale $10^5$--$10^8$\,yr \citep{vigmel09b}. 

A coherent search for 78 known radio pulsars was performed on S3 and S4 LIGO and GEO 600 data. Upper limits on the ellipticities of these pulsars were obtained, the smallest being $\epsilon \leq 7.1 \times 10^{-7}$ for PSR J2124--3358 \citep{lscpulsars07}. More recently, a coherent search for 116 known pulsars was carried out using data from both the LIGO and Virgo detectors, placing an upper limit of $\epsilon < 7.0 \times 10^{-8}$ for PSR J2124$-$3358.

The youngest isolated neutron star accessible to LIGO probably resides in the supernova remnant SNR 1987A. The coincident detection of neutrino bursts from the supernova by detectors all over the world confirmed the core-collapse event, strongly indicating the formation of a neutron star \citep{aglietta87, hirata87, bionta87, bahcall87}.\footnote{An unconfirmed correlation was also reported between data taken by the Mont Blanc and Kamioka neutrino detectors and gravitational wave detectors in Maryland and Rome \citep{amaldi89}. Taken at face value, these observations are consistent with a weak neutrino pulsar operating briefly during the core-collapse event. However, a serious flaw in the original analysis was found by \citet{dickson95}, whose reanalysis led them to conclude that the correlations were not physically significant.}
 Constraints have been placed on the magnetic field strength, spin period, and other birth properties of the putative neutron star \citep{michel94, ogelman04}; see Section \ref{sn1987Adetails} for details. However, searches for a pulsar in SNR 1987A have yielded no confirmed sightings; upper limits on its luminosity have been placed in the radio, optical and X-ray bands \citep{percival95, burrows00, manchester07}. An unconfirmed detection of a transitory 467.5 Hz optical/infra-red pulsation in SNR 1987A was reported by \citet{middleditch00}. 

 The likely existence of a young neutron star in SNR 1987A makes it a good target for gravitational wave searches \citep{piran88, nakamura89}.
 A coherent matched filtering search was carried out in 2003 with the TAMA 300 detector, searching $1.2 \times 10^3$ hours of data from its first science run over a 1-Hz band centered on 934.9\,Hz, assuming a spin-down range of (2--3)$\times 10^{-10}$\,Hz s$^{-1}$. The search yielded an upper limit on the wave strain of $5 \times 10^{-23}$ \citep{soida03}. An earlier matched filtering search was conducted using $10^2$ hours of data taken in 1989 by the Garching prototype laser interferometer. The latter search was carried out over 4-Hz bands near 2\,kHz and 4\,kHz, did not include any spin-down parameters, and yielded an upper limit of $9 \times 10^{-21}$ on the wave strain \citep{niebauer93}.

There are two main types of continuous-wave LIGO searches: coherent and semi-coherent. The former demand phase coherence between the signal and search template over the entire time series. Although sensitive, they are restricted to small observation times and parameter ranges as they are computationally intensive. Semi-coherent searches break the full time series into many small chunks, analyse each chunk coherently, then sum the results incoherently, trading off sensitivity for computational load.
 \citet{santostasi03} discussed the detectability of gravitational waves from SNR 1987A, estimating that a coherent search based on the \citet{middleditch00} spin parameters requires 30 days of integration time and at least $10^{19}$ search templates covering just the frequency and its first derivative. 
In reality, the task is even more daunting, because such a young object spins down so rapidly, that five or six higher-order frequency derivatives must be searched in order to accurately track the gravitational wave phase. A Bayesian Markov Chain Monte Carlo method was proposed as an alternative to cover the parameter space efficiently \citep{umstaetter04, umstaetter08}. As yet, though, SNR 1987A has not been considered a feasible search target, because even Monte Carlo methods are too arduous. In this paper, we show how to reduce the search space dramatically by assuming an astrophysically motivated phase model.

In this paper, we discuss how to use a cross-correlation algorithm to search for periodic gravitational waves from a neutron star in SNR 1987A. The search is semi-coherent \citep{dhurandhar08}. The signal-to-noise ratio is enhanced by cross-correlating two data sets separated by an adjustable time lag, or two simultaneous data sets from different interferometers, thereby nullifying short-term timing noise (e.g. from rotational glitches). This is a modification of the method used in searches for a cosmological stochastic background \citep{lscstochastic07, lscstochastic09} and for the low-mass X-ray binary Sco X-1 \citep{lscscox107}. In Section \ref{sn1987Adetails}, we review the properties of SNR 1987A and its putative neutron star.
Section \ref{sec:crosscorr} briefly describes the cross-correlation algorithm and the data format. We estimate the theoretical sensitivity of the search in Section \ref{sec:ccsensitivity}.
 Section \ref{searchparams} describes an astrophysical model, which expresses the gravitational wave phase in terms of the initial spin, ellipticity, magnetic field, and electromagnetic braking index of the neutron star. 
We calculate the semi-coherent phase metric and the number of templates required for the search in the context of the astrophysical phase model. Given the computational resources available to us, we derive upper limits on the gravitational wave strain, ellipticity and magnetic field which can be placed on a neutron star in SNR 1987A with a cross-correlation search. Finally, Section \ref{sec:ccconclusion} summarises the results.


\section{A young neutron star in SNR 1987A}
\label{sn1987Adetails}

SNR 1987A is the remnant of a Type II core-collapse supernova which occurred in February 1987, 51.4\,kpc away in the Large Magellanic Cloud ($\alpha$ = 5h 35m 28.03s, $\delta$ = $-$69$^\circ$ 16$^\prime$ 11.79$^{\prime\prime}$) \citetext{see reviews by \citealp{panagia08} and in \citealp{immler20yrs07}}. Its progenitor was the blue supergiant Sk 1  \citep{panagia87, gilmozzi87, barkat88, woosley02}. The color of the progenitor, as well as the origin of the complex three-ring nebula in the remnant, are still unexplained. Detailed simulations of the evolutionary history of Sk 1, performed by \citet{podsiadlowski07}, support the theory that two massive stars merged to form an oversized $20 M_\odot$ red supergiant $2 \times 10^5$ years before the supernova, which eventually shrank as its envelope evaporated \citep[e.g.][]{podsiadlowski89, podsiadlowski90}. An alternative theory suggests that Sk 1 was instead a single 18--20\,$M_\odot$ red supergiant which evolved into a blue supergiant via wind-driven mass loss \citep[e.g.][]{woosley88, saio88, sugerman05}.

There is strong evidence for the existence of a neutron star in SNR 1987A. The progenitor mass range required to produce Type II supernovae, 10--25\,$M_\odot$, which includes the above evolutionary scenarios, is the same range required to produce neutron star remnants \citep{woosley02, heger03}.  The secure neutrino detections mentioned in Section \ref{sec:ccintro} support this conclusion. Although there have been no confirmed pulsar detections, numerous searches have placed upper limits on the flux and luminosity at radio \citep[$< 115$ $\mu$Jy at 1390 MHz, ][]{manchester07}, optical/near-UV \citep[$< 8 \times 10^{33}$\,ergs s$^{-1}$, ][]{graves05}, and soft X-ray \citep[$< 2.3 \times 10^{34}$ erg s$^{-1}$,][]{burrows00} wavelengths. \citet{middleditch00} reported finding an optical pulsar in SNR 1987A with a frequency of 467.5\,Hz, modulated sinusoidally with a $\sim$ 1-ks period, consistent with precession for an ellipticity of $\epsilon \sim 10^{-6}$. However, the pulsations were reported to have disappeared after 1996 \citep{middleditch00} and were never confirmed independently.

 There are several possible reasons why a pulsar in SNR 1987A has not yet been detected. If its spin period is greater than 0.1\,s, it would not be bright enough to be detectable in the optical band \citep{pacini87, manchester07}. If the radio emission is incoherent or the emission region is patchy, the pulses may have been missed, even if the beam width is as wide as is typical for young pulsars \citep{manchester07}. \citet{shternin08} argued that, although the neutron star's theoretical X-ray luminosity exceeds the observational upper limits by a factor of 20--100, the current upper limits still allow for concealment behind an opaque shell formed by fallback \citep{woosley95}. However, simulations by \citet{fryer99} suggest that, once fallback ceases, the accreted material cools, leaving no obscuring atmosphere. 

Another possible reason why a pulsar has not yet been detected is that its magnetic field is too weak.
The weak-field theory is supported by theoretical models, in which the field grows only after the neutron star is formed and can take up to $10^3$ years to develop \citep[e.g.][]{blandford88, reisenegger03}. A growth model for SNR 1987A was proposed by \citet{michel94}, in which the magnetic field of a millisecond pulsar intensifies from $10^{10}$\,G  at birth to $\sim 10^{12}$\,G after several hundred years (exponential and linear growth were considered, yielding growth times of $\sim$ 0.3--0.7 kyr), before the pulsar has time to spin down significantly. In an alternative model, the neutron star is born with a strong magnetic field, which is amplified during the first few seconds of its life by dynamo action \citep[e.g.][]{duncan92, bonanno05}. Assuming this model, measurements of the known spin periods of isolated radio pulsars imply a distribution of birth magnetic field strengths between $10^{12}$G and $10^{13}$G \citep{arzoumanian02, faucher06}. Several birth scenarios for the pulsar in SNR 1987A were considered by \citet{ogelman04} in this context, who concluded that the maximum magnetic dipole moment is $< 1.1 \times 10^{26}$\,G cm$^3$, $2.5 \times 10^{28}$\,G cm$^3$, and $2.5 \times 10^{30}$\,G cm$^3$ for birth periods of 2\,ms, 30\,ms, and 0.3\,s respectively. However, the dynamo model also accommodates a magnetar in SNR 1987A, with magnetic dipole moment $> 2.4 \times 10^{34}$\,G cm$^3$, regardless of the initial spin period \citep{ogelman04}.

Estimates of the birth spin of the pulsar in SNR 1987A are more uncertain. Simulations of the bounce and post-bounce phases of core collapse were performed by \citet{ott06} to determine the correlation between progenitor properties and birth spin. These authors found proto-neutron star spin periods of between 4.7--140 ms, proportional to the progenitor's spin period. A Monte Carlo population synthesis study using known velocity distributions \citep{arzoumanian02} favoured shorter millisecond periods, but a similar population study by \citet{faucher06} argued that the birth spin periods could be as high as several hundred milliseconds. Faint, non-pulsed X-ray emission from SNR 1987A was first observed four months after the supernova and decreased steadily in 1989 \citep{dotani87, inoue91}, leading to the suggestion that a neutron star could be powering a plerion that is partially obscured by a fragmented supernova envelope. \citet{bandiera88} modelled the X-ray spectrum from a nebula containing a central pulsar, with a magnetic field of $10^{12}$\,G and an expansion rate of $5 \times 10^8$\,cm s$^{-1}$. The authors found a fit to the SNR 1987A data for a pulsar spin period of 18\,ms.

\section{The cross-correlation algorithm}
\label{sec:crosscorr}

In this section, we briefly summarise the cross-correlation method described in \citet{dhurandhar08}, a semi-coherent search algorithm designed specifically to search for continuous-wave gravitational radiation. It operates on Short Fourier Transforms (SFTs) of data segments of length $\Delta T = 30$ min, whose duration is chosen to minimise the Doppler effects due to Earth's rotation. In each SFT, the $k$th frequency bin corresponds to the frequency $\nu_k = k/\Delta T$ for $0 \leq k \leq N/2$ and $\nu_k = (k-N)/\Delta T$ for $N/2 \leq k \leq N-1$, where $N$ is the total number of frequency bins in the SFT.  

The output $x(t)$ of a detector is the sum of the instantaneous noise, $n(t)$, and the gravitational wave signal, $h(t)$.  The noise is assumed to be zero mean, stationary, and Gaussian. Its power is characterised by $S_n(\nu)$, the single-sided power spectral density (i.e. the frequency-dependent noise floor) in the following way:
\begin{equation}
\label{eq:noisesq}
\langle \tilde{n}(\nu)^* \tilde{n}(\nu')\rangle = \frac{1}{2} S_n(\nu) \delta (\nu - \nu'),
\end{equation} 
where $^*$ denotes complex conjugation. Therefore, in the low signal limit ($\lvert h(t) \rvert \ll \lvert n(t) \rvert$), the power in the $k$-th frequency bin of SFT $I$ can be approximated by
\begin{equation}
\langle \lvert \tilde{x}_{k, I} \rvert ^2 \rangle \approx \frac{\Delta T}{2} S_n(\nu_k),
\end{equation}
where we apply the finite time approximation to the delta function in (\ref{eq:noisesq}), i.e. $\delta_{\Delta T} (\nu) = \sin (\pi \nu \Delta T)/(\pi \nu) \approx \Delta T$.

In the cross-correlation algorithm, SFTs are paired according to some criterion (e.g. time lag or interferometer combination) and multiplied to form the raw cross-correlation variable
\begin{equation}
\label{eq:yalpha}
\mathcal{Y}_{k, IJ} = \frac{\tilde{x}^*_{k_I, I} \tilde{x}_{k_J,J}}{ (\Delta T)^2},
\end{equation}
where $I$ and $J$ index the SFTs in the pair. The gravitational wave signal is assumed to be concentrated in a single frequency bin in each SFT (because $\Delta T \ll \nu/\dot{\nu}$ due to sidereal or intrinsic effects), whose index is denoted by $k_I$ or $k_J$. The frequency bins in the two SFTs are not necessarily the same; they are related by the time lag between the pair and between interferometers, as well as spin-down and Doppler effects.  For an isolated source, the instantaneous frequency at time $t$ is given by
\begin{equation}
\label{eq:freq}
\nu(t) = \hat{\nu}(t) + \hat{\nu}(t) \frac{\mathbf{v}\cdot\mathbf{n}}{c},
\end{equation}
where $\hat{\nu}(t)$ is the instantaneous frequency in the rest frame of the source, $\mathbf{v}$ is the detector velocity relative to the source, $\mathbf{n}$ is the position vector pointing from the detector to the source, and $c$ is the speed of light. 
The instantaneous signal frequencies in SFTs $I$ and $J$, $\nu_I$ and $\nu_J$, are calculated at the times corresponding to the midpoints of the SFTs, $T_{I}$ and $T_J$. The frequency bin $k_J$ is therefore shifted from $k_I$ by an amount $\Delta T \delta \nu_{IJ}$, with $\delta \nu_{IJ} = \nu_J - \nu_I$ \citep{dhurandhar08}. For convenience, we now drop the subscripts $k_I$ and $k_J$.

In the low signal limit, $\mathcal{Y}_{IJ}$ is a random, complex variable. The cross-correlation statistic comprises a weighted sum of $\mathcal{Y}_{IJ}$ over all pairs $IJ$. $\mathcal{Y}_{IJ}$ has variance $\sigma^2_{IJ} = S_n^{(I)}(\nu_I) S_n^{(J)}(\nu_J)/ (4 \Delta T^2)$, where $S_n^{(I)}(\nu_I)$ is the power spectral density of SFT $I$ at frequency $\nu_I$, and $S_n^{(J)}(\nu_J)$ is the power spectral density of SFT $J$ at frequency $\nu_J$. 

The parameters describing the amplitude and the phase of the signal are contained within the signal cross-correlation function $\tilde{\mathcal{G}}_{IJ}$, defined as
\begin{eqnarray}
\nonumber \tilde{\mathcal{G}}_{IJ} &=& \frac{1}{4} e^{-i\Delta\Phi_{IJ}} e^{-i \pi \Delta T[\nu_I(T_I) - \nu_J (T_J)]} \left[F_{I+} F_{J+} \mathcal{A}_+^2   \right. \\
\label{eqngij} && \left. + F_{I\times} F_{J\times} \mathcal{A}^2_{\times} - i(F_{I+} F_{J\times} - F_{I\times}F_{J+}) \mathcal{A}_+ \mathcal{A}_{\times}\right],
\end{eqnarray}
with $\Delta \Phi_{IJ} = \Phi_I(T_I) - \Phi_J(T_J)$. $\Phi_I(T_I)$ and $\nu_I(T_I)$ are the phase and frequency at time $T_I$, whereas $\Phi_J(T_J)$ and $\nu_J(T_J)$ are evaluated at time $T_J$. Note that there is an error in equation (3.10) of \citet{dhurandhar08}, which omits the factor of $e^{-i \pi \Delta T [\nu_I (T_I) - \nu_J (T_J)]}$ arising from the choice of time origin of the Fourier transforms.
The phase factors are determined by the astrophysical phase model described in Section \ref{searchparams}. 

The terms in square brackets in (\ref{eqngij}) depend on the polarization angle $\psi$, and the inclination angle $\iota$ between $\mathbf{n}$ and the rotation axis of the pulsar, in the following way:
\begin{eqnarray}
\label{eq:curlyaplus} \mathcal{A}_+ &=& \frac{1 +  \cos^2 \iota}{2} ,\\
\mathcal{A}_\times &=& \cos \iota ,\\
F_+(t; \mathbf{n}, \psi) &=& a(t; \mathbf{n}) \cos 2\psi + b(t; \mathbf{n}) \sin 2\psi,\\
\label{eq:ftimes} F_\times(t; \mathbf{n}, \psi) &=& b(t; \mathbf{n}) \cos 2\psi - a(t; \mathbf{n}) \sin 2\psi,
\end{eqnarray}
where $a(t;\mathbf{n})$ and $b(t; \mathbf{n})$ are the detector response functions for a given sky position, and are defined in equations (12) and (13) of \citet{jara98}. A geometrical definition is also given in \citet{prixwhelan07}. The gravitational wave strain tensor is
\begin{equation}
\overleftrightarrow{h}(t) = h_0 \mathcal{A}_+ \cos \Phi(t) \overleftrightarrow{e}_+ + h_0 \mathcal{A}_\times \sin \Phi(t) \overleftrightarrow{e}_\times
\end{equation} 
where $h_0$ is the gravitational wave strain, and $\overleftrightarrow{e}_{+, \times}$ are the basis tensors for the + and $\times$ polarizations in the transverse-traceless gauge.

In principle, one should search over the unknowns $\cos \iota$ and $\psi$, but this adds to the already sizeable computational burden. Accordingly, it is customary to average over $\cos \iota$ and $\psi$ when computing $\tilde{\mathcal{G}}_{IJ}$, with
\begin{equation}
\label{eq:galphave}
\langle \tilde{\mathcal{G}}_{IJ}\rangle_{\cos \iota, \psi} = \frac{1}{10} \exp^{-i \Delta \Phi_\text{IJ}} e^{-i \pi \Delta T[\nu_I(T_I) - \nu_J (T_J)]} (a_I a_J + b_I b_J),
\end{equation}
where $a_{I, J} = a (T_{I,J}; \mathbf{n})$ and $b_{I, J} = b (T_{I, J}; \mathbf{n})$. Once the first-pass search is complete, a follow-up search on any promising candidates can then be performed, which searches explicitly over $\cos \iota$ and $\psi$. Preliminary Monte Carlo tests indicate that the detection statistic resulting from (\ref{eq:galphave}) is approximately 10$-$15\% smaller than if the exact $\cos \iota$ and $\psi$ values are used.


 The cross-correlation detection statistic is a weighted sum of $\mathcal{Y}_{IJ}$ over SFT pairs. The number of pairs which can be summed over are limited by the available computational power. We discuss the computational costs of running the search in Section \ref{sec:templates}. The cross-correlation detection statistic is given by
\begin{equation}
\label{eq:rho}
\rho = \Sigma_{IJ} (u_{IJ} \mathcal{Y}_{IJ} + u^*_{IJ} \mathcal{Y}^*_{IJ}),
\end{equation}
where the weights are defined by
\begin{equation}
\label{eq:ualpha}
u_{IJ} = \frac{\tilde{\mathcal{G}}_{IJ}^*}{\sigma_{IJ}^2}.
\end{equation}
For each frequency and sky position that is searched, we obtain \textit{one} real value of $\rho$, which is a sum of the Fourier power from all the pairs. Ignoring self-correlations (i.e. no SFT is paired with itself), the mean of $\rho$ is given by $\mu_\rho = h_0^2 \sum_{IJ} \lvert \tilde{\mathcal{G}}_{IJ} \rvert^2/ \sigma_{IJ}^2$. In the low signal limit, the variance of $\rho$ is $\sigma^2_\rho = 2 \Sigma_{IJ} \lvert \tilde{\mathcal{G}}_{IJ}\rvert^2 /\sigma^2_{IJ}$. In the presence of a strong signal, and if self-correlations are included, $\mu_\rho$ and $\sigma^2_\rho$ scale as $h_0^2$ \citep{dhurandhar08}. 
\section{Sensitivity}
\label{sec:ccsensitivity}

\subsection{Detection threshold}
\label{sec:detthresh}
Detection candidates are selected if they exceed a threshold value, $\rho_\text{th}$.  For a given false alarm rate $F_a$, this threshold is given by \citep{dhurandhar08}
\begin{equation}
\label{eq:rhothresh} \rho_\text{th} = 2^{1/2} \sigma_\rho \text{erfc}^{-1} (2 F_a/N),
\end{equation}
where erfc is the complementary error function, and $N$ is the number of search templates used.
In the presence of a signal, the detection rate for events with $\rho > \rho_\text{th}$ is given by
\begin{equation}
\gamma = \frac{1}{2} \text{erfc}\left(\frac{\rho_\text{th} - \mu_\rho}{ \sqrt{2} \sigma_\rho}\right).
\end{equation}
As $\mu_\rho \propto h_0^2$, one can calculate the lowest gravitational wave strain that is detectable by the search to be \citep{dhurandhar08} 
\begin{equation}
\label{eq:hth}
h_\text{th}(\nu) = \frac{\mathcal{S}^{1/2}}{\sqrt{2} \langle \lvert\tilde{\mathcal{G}}_{IJ}\rvert^2 \rangle^{1/4} N_{\text{pairs}}^{1/4}} \left[\frac{S_n(\nu)}{\Delta T}\right]^{1/2}.
\end{equation}
In (\ref{eq:hth}), we define $\mathcal{S} =$ erfc$^{-1} (2 F_a) + $erfc$^{-1} (2 F_d)$, $F_d$ is the false dismissal rate, $\langle \lvert \tilde{\mathcal{G}}_{IJ}\rvert^2 \rangle$ is the mean-square of the signal cross-correlation function defined in (\ref{eqngij}), $N_{\text{pairs}}$ is the number of SFT pairs, and $S_n (\nu)$ is the single-sided power spectral density of the interferometers (assumed to be identical).

One can estimate  $\langle |\mathcal{G}_{IJ}|^2 \rangle^{1/4}$ theoretically for the special case where $T_I = T_J$ and $\tilde{\mathcal{G}}_{IJ}$ is averaged over $\cos \iota$, $\psi$, and sidereal time. In this case, the primary contribution to $\Delta \Phi_{IJ}$ is the term $[\mathbf{r}(T_I) - \mathbf{r}(T_J)] \cdot \mathbf{n} / c$, where $\mathbf{r}(t)$ is the position of the detector at time $t$ in the frame of the solar system barycentre. Under these assumptions, equation (\ref{eq:galphave}) can be expressed in terms of the overlap reduction function \citep{whelan06}, which depends only on $\nu, \alpha,$ and $\delta$. 
For SNR 1987A, we have $(\alpha, \delta)$ = (1.46375 rad, $-$1.20899 rad), and hence $\langle |\tilde{\mathcal{G}}_{IJ}|^2 \rangle^{-1/4} = 4.6882$. Assuming $F_a = F_d = 0.1$, $\Delta T = 1800$\,s, and $N_\text{pairs} = 10^5$ (approximately 1 year of SFTs), equation (\ref{eq:hth}) gives
\begin{equation}
\label{hthreshold}
h_\text{th}(\nu) = 5.92 \times 10^{-3} \left[\frac{S_n(\nu)}{\text{Hz}^{-1}}\right]^{1/2}.
\end{equation} 

\begin{figure*}
\centering
\scalebox{0.65}{\includegraphics{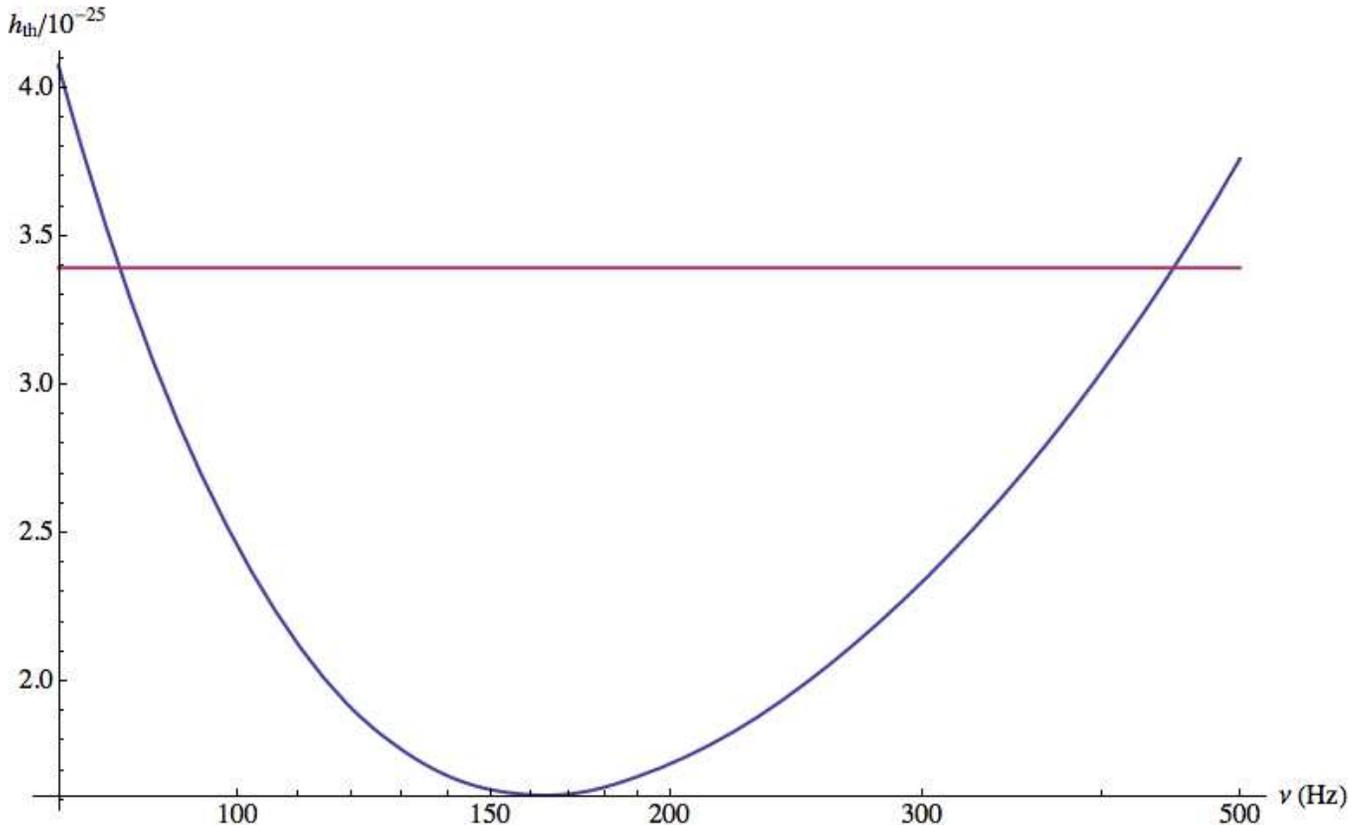}}
\caption[Theoretical sensitivity of the cross-correlation search for SNR 1987A]{Theoretical sensitivity of the cross-correlation search for SNR 1987A as a function of gravitational wave frequency (blue curve), assuming the initial LIGO detector power spectral density, a false alarm rate of 0.1, and a false dismissal rate of 0.1. The blue curve shows the theoretical sensitivity for the special case where the search uses $10^5$ pairs of time-coincident 30-minute SFTs, and averages over inclination angle, polarization angle, and sidereal time (see discussion in Section \ref{sec:detthresh}). The horizontal red line shows the indirect, age-based limit assuming $\nu \ll \nu_b$ (see discussion in Section \ref{sec:agebased}).}
\label{hth}
\end{figure*}

Figure \ref{hth} is a graph of $h_\text{th}$ as a function of $\nu$. The values of $S_n(\nu)$ are based on LIGO's S5 noise characteristics.\footnote{Available at http://www.ligo.caltech.edu/\~jzweizig/distribution/LSC\_Data} The S5 run began in November 2005 and accumulated a year's worth of triple coincidence data.
For a signal from SNR 1987A to be detectable, we must have $h_\text{th} \leq h_0$.

\subsection{Minimum ellipticity and indirect, age-based limit}
\label{sec:agebased}
 The deformation of a neutron star is parameterised by its ellipticity $\epsilon$. The gravitational wave strain at Earth emitted by a biaxial neutron star is
\begin{equation}
\label{hstrain}
h_0 = \frac{4 \pi^2G}{c^4} \frac{I \epsilon \nu^2}{D}
\end{equation}
where $G$ is Newton's gravitational constant, $c$ is the speed of light, $I$ is the moment of inertia, $D$ is the distance to the source, and $\nu$ is the gravitational wave frequency, assumed to be twice the spin frequency \citep{jara98}.

An upper limit on $h_0$ can be derived from existing electromagnetic data by assuming all the observed spin down comes from the gravitational wave torque, i.e. the observed frequency derivative $\dot{\nu}$ satisfies $\dot{\nu} = -(32 \pi^4 G \epsilon^2 I \nu^5)/(5 c^5)$ \citep{wette08}. Combining this with (\ref{hstrain}) to eliminate $\epsilon$ gives
\begin{equation}
\label{eq:h0limit}
h_0 \leq \frac{1}{D} \left(\frac{5 G I \lvert \dot{\nu} \rvert}{2 c^3 \nu}\right)^{1/2} .
\end{equation}
Hence, for SNR 1987A to be detectable (i.e. $h_\text{th} \leq h_0$), we require
\begin{eqnarray}
\nonumber h_\text{th} (\nu) &\leq& 1.66 \times 10^{-20} \\
\label{eq:hthnudot} && \times \left(\frac{I}{10^{38}\rm{kg\,m^2/s}}\right)^{1/2} \left(\frac{\lvert\dot{\nu}\rvert}{\nu}\right)^{1/2} \left(\frac{D}{51.4 \rm{kpc}}\right)^{-1}
\end{eqnarray}

Unfortunately, without having observed any pulsations from SNR 1987A, it is impossible to determine $\nu$ or $\lvert \dot{\nu} \rvert$ a priori. Instead, we note that $\dot{\nu}$ can be re-expressed in terms of the characteristic age of the source, $\tau_c = \nu/(4 \lvert \dot{\nu} \rvert)$, assuming that $\nu$ today is much less than $\nu$ at birth. The factor 4 arises if one assumes the gravitational radiation dominates electromagnetic spin down, in order to remain consistent with (\ref{eq:h0limit}); in reality, electromagnetic spin down is expected to dominate, with $\tau_c = \nu/(2 \lvert \dot{\nu} \rvert)$. 
Equation (\ref{eq:hthnudot}) then reduces to
\begin{equation}
\label{eq:hthreduced}
h_\text{th} (\nu) \leq 3.39 \times 10^{-25} \left(\frac{\tau_c}{19\,\text{yr}}\right)^{-1/2} \left(\frac{D}{51.4 \rm{kpc}}\right)^{-1}.
\end{equation}
The right-hand side of (\ref{eq:hthreduced}) is graphed as a horizontal red line in Figure \ref{hth}. The detectability condition (\ref{eq:hthreduced}) is then satisfied for spins in the range $75\,\text{Hz} \lesssim \nu \lesssim 450\,\text{Hz}$.  Note that we have chosen $\tau_c = 19$\,yr, the age of SNR 1987A in 2006 when the S5 search began.

It is important to note here that the assumption that $\nu$ is currently much less than at birth is likely \textit{untrue} for the object in SNR 1987A, as it is so young. Hence, the indirect, age-based limit in equation (\ref{eq:hthreduced}) and the horizontal line in Figure \ref{hth} are only indicative of the expected gravitational-wave emission strength (in fact, they are upper limits). Exact calculations of $\nu$ and $\dot{\nu}$ are performed in Section \ref{sec:historical}.

\section{An astrophysical model for the gravitational wave phase}
\label{searchparams}

All continuous wave searches to date have used the standard model for the gravitational wave phase, described in terms of a Taylor expansion involving spin frequency derivatives \citep{jara98}. For a young object like SNR 1987A, which spins down rapidly, it is not computationally feasible to search over the six or more frequency derivatives typically needed to track the phase accurately.
In this section, we present an alternative model for the gravitational wave phase, stated in terms of astrophysical parameters (i.e. the magnetic field strength and the neutron star ellipticity) instead of spin frequency derivatives. It tracks the phase exactly using four parameters, under the restrictive assumption (justified further below) that the braking index is constant.

The phase of a slowly evolving gravitational wave signal, 
\begin{equation}
\label{eq:phit}
\Phi(t) = \Phi(t_0) + 2\pi \int_{t_0}^t\,dt\,\nu(t),
\end{equation}
can be approximated by the Taylor expansion \citep{jara98}
\begin{equation}
\label{eq:phitaylor}
\Phi(t) = \Phi(t_0) + 2\pi \sum_{k=0}^s \nu^{(k)} \frac{t^{k+1}}{(k+1)!} + \frac{2\pi \mathbf{n}\cdot\mathbf{r}(t_0)}{c} \sum_{k=0}^s \nu^{(k)} \frac{t^k}{k!}
\end{equation}
where $\nu^{(k)}$ is the $k$-th derivative of the gravitational wave frequency at time $t_0$, and $s$ is the number of spin-down parameters required to achieve a given accuracy. 
The computational cost of using (\ref{eq:phitaylor}) is substantial for rapidly decelerating objects. For a maximum allowable phase error of one cycle, the maximum bin size in the $k$-th derivative is $\nu^{(k)}$ is $\Delta \nu^{(k)} = (k + 1)!/\Tlag^{k+1}$, implying $N_{k} \approx \nu^{(k)}/ \Delta \nu^{(k)}$ templates in that derivative and $N_\text{total} = \prod^s_{k=0} N_k$ templates overall. We discuss this matter further in Section \ref{sec:templates}.

To improve on the above situation, we recognize that $\dot{\nu}$ for an isolated neutron star is the sum of gravitational-wave and electromagnetic torque contributions:
\begin{eqnarray}
\dot{\nu} &=& -\frac{32 \pi^4 G \epsilon^2 I \nu^5}{5 c^5} - \frac{2 \pi^3 R_{\star}^6 B^2 \nu^n}{3 \mu_0 I c^3} \left(\frac{\pi R_{\star}}{c}\right)^{n-3}\\
\label{eq:numodel} &=& - Q_1' \nu^5 - Q_2' \nu^n,
\end{eqnarray}
where $R_{\star}$ is the neutron star radius, $B$ is the polar magnetic field, $n$ is the electromagnetic braking index \citep[theoretically equal to 3, but could be as low as 1.8;][]{melatos97, palomba05}. Assuming that the electromagnetic torque is proportional to a power of $\nu$, then $\nu$ must enter the torque in the combination $R_\star \nu/c$, (i.e. the ratio of $R_\star$ to the characteristic lever arm, the light cylinder distance, $c/2 \pi \nu$) on dimensional grounds.
In terms of an arbitrary reference frequency, $\nu_{\rm{ref}}$, we write $\dot{\nu} = - Q_1 \left(\nu/\nu_{\rm{ref}}\right)^5 - Q_2 \left(\nu/\nu_{\rm{ref}}\right)^n$, with $Q_1 = Q_1' \nu_{\rm{ref}}^5$ and $Q_2 = Q_2' \nu_{\rm{ref}}^n$. Throughout this paper, we set $\nu_\text{ref} = 1$\,Hz for simplicity.

There may, of course, be other torques acting on a newly born neutron star. For example, nonlinear r-mode instabilities can emit a significant amount of gravitational radiation under certain conditions \citep{owen98}. If there is a rapidly rotating pulsar with $B \leq 10^{11}$\,G in SNR 1987A, its instability time scale (27 years) would exceed its age, and the gravitational radiation from the instabilities alone should be detectable by Advanced LIGO \citep{brink04, bondarescu09}. However, for the purposes of our search, we assume that the spin down is described by (\ref{eq:numodel}).
An equally serious issue is that $n$ may change over the 1\,yr integration period, although in (\ref{eq:numodel}), we assume that $n$ is constant. Young pulsars have $n < 3$, and it can be argued that $n$ approaches 3 over the spin-down time-scale \citep{melatos97}. In this search, we maintain the assumption of constant $n$. However, it is possible to extend (\ref{eq:numodel}) to include time-dependent $n$ in future searches. We aim, in the first instance, to exclude the simplest astrophysical model while recognizing that it covers only a small fraction of the total parameter space.

When implementing the search, instead of stepping through a grid of frequency derivatives, we search instead over $\nu, Q_1, Q_2$, and $n$. This reduces the number of parameters and allows one to track the phase more accurately for a given computational cost, as errors stemming from incorrect choices of $(\nu, Q_1, Q_2, n)$ grow more slowly with observation time than errors stemming from higher-order frequency derivatives. The improvement is quantified in Section \ref{sec:templates}. We note that the search targets a source with a known position, hence in our estimates we consider only a single sky position. 


\subsection{Historical spin down}
\label{sec:historical}
We can use the possible spin histories of a source like SNR 1987A with a known age to constrain the invisible values of $(\nu, Q_1, Q_2, n)$ today and hence the maximum amount of phase evolution to be expected during a LIGO integration.

There are two ways of estimating $\nu$ and $\dot{\nu}$ for a source whose age is known. In the simplest situation, where the current spin frequency $\nu$ is much smaller than the \textit{birth} frequency $\nu_b$, the characteristic age $\tau_c \approx - \nu / \left[\left(\langle n \rangle - 1 \right) \dot{\nu} \right]$ closely approximates the true age irrespective of $\nu_b$, where $\langle n \rangle$ is the mean braking index, averaged over the time since birth. Under these conditions, a source with unknown $\nu$ and $\dot{\nu}$ lies on a line of slope $-\tau_c \left(\langle n \rangle - 1 \right)$ in the $\nu$-$\dot{\nu}$ plane. However, as discussed in Section \ref{sec:agebased}, this is not necessarily true for SNR 1987A, which was only 19 years old at the start of the S5 search.
In order to calculate $\nu$ and $\dot{\nu}$ exactly without using the characteristic age approximation, one must integrate (\ref{eq:numodel}) over the lifetime of the source.
Accordingly, we adopt this approach and map out the regions in the $\nu$-$\dot{\nu}$ plane which can be reached from $\nu_b$ by electromagnetic-plus-gravitational-wave spin down and physically sensible choices of $\epsilon, B$ and $n$. 

\begin{figure*}
\centering
\scalebox{0.65}{\includegraphics{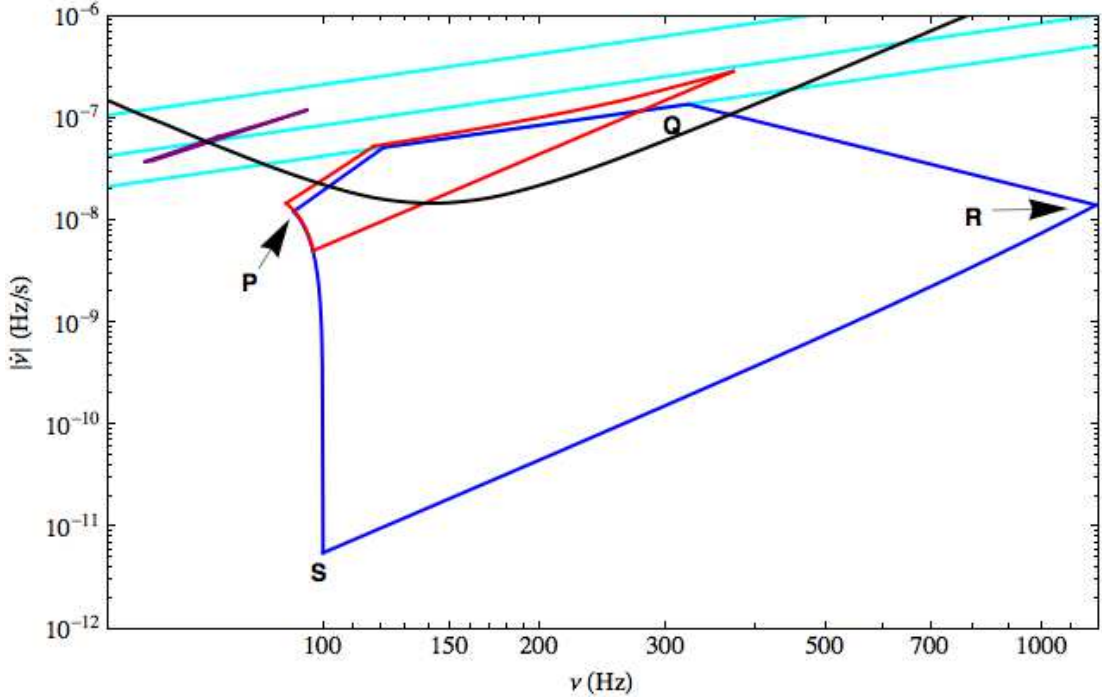}}
\caption[Gravitational wave and electromagnetic dominated regions in the $\nu, \dot{\nu}$ plane.]{Final states ($\nu$, $\dot{\nu}$) calculated from equation (\ref{eq:numodel}) on the $\nu$-$\lvert \dot{\nu} \rvert$ plane for a range of ellipticities ($10^{-6} \leq \epsilon \leq 10^{-3}$), and birth spin frequencies ($0.10$\,kHz $\leq \nu_b \leq 1.2$\,kHz), and for a 19 yr old pulsar. The blue lines surround the region where the $Q_1$ term dominates ($B \leq 10^{11.5}$\,G, all $n$), the red lines surround the region where the $Q_2$ term dominates ($B = 10^{13}\,\text{G}, n = 3$), and the purple lines surround the region where the $Q_2$ term dominates ($B = 10^{13}\,\text{G}, n = 2.3$). The black curve shows the theoretical search sensitivity from solving equation (\ref{hthreshold}). The $\nu \ll \nu_b$ age limits are shown in cyan for $\langle n \rangle = 1.8$ (top), $\langle n \rangle = 3$ (middle) and $ \langle n \rangle = 5$ (bottom). }
\label{ffdot_large}
\end{figure*}

Figure \ref{ffdot_large} shows the range of possible $\nu$ and $\dot{\nu}$ values at $t = 19$\,yr obtained by solving (\ref{eq:numodel}) for $10^{-6} \leq \epsilon \leq 10^{-3}$, $10^{11.5} \leq B \leq 10^{13}\,\rm{G}$ and $0.1\,\rm{kHz} \leq \nu_b \leq 1.2\,\rm{kHz}$. 
For reference, we plot the search sensitivity (black curve in the $\nu$-$\dot{\nu}$ plane) obtained from (\ref{hthreshold}). According to (\ref{hthreshold}), the search is only sensitive to combinations of $\nu$ and $\dot{\nu}$ above the black line. The conservative limits set by the characteristic age approximation are plotted as cyan lines. The lines correspond to $\langle n \rangle = 1.8$ (top), $\langle n \rangle = 3$ (middle) and $\langle n \rangle = 5$ (bottom). For a given value of $\langle n \rangle$, an object lies on the line for $\nu \ll \nu_b$, and below the lines for $\nu \lesssim \nu_b$, but never above the line. 
 
The blue, red and purple boxes contain combinations of ($\nu$, $\dot{\nu}$) that can be reached for various choices of $\epsilon$, $B$, $n$, and $\nu_b$. The blue box covers the region in which $B \leq 10^{11.5}$\,G and $n = 3$, and the gravitational wave torque ($Q_1$) dominates, i.e. $\dot{\nu}_\text{GW} \gg \dot{\nu}_\text{EM}$, where the subscripts EM and GW denote the electromagnetic and gravitational wave components of the spin down respectively. The red box covers the region in which the electromagnetic torque ($Q_2$) dominates, with $B = 10^{13}$\,G and $n = 3$. The purple box also shows a region in which the $Q_2$ term dominates, where we have chosen $B = 10^{13}$\,G and $n = 2.3$. As a rule of thumb, $\epsilon$ determines the size of the box along the $\nu$-axis, and $\nu_b$ determines the size of the box along the $\dot{\nu}$-axis.

Let us first investigate what happens to the blue box when we vary the minimum and maximum ellipticity, $\epsilon_\text{min}$ and $\epsilon_\text{max}$. The $Q_1$ term dominates in the region bounded by the blue box. The absolute value of the $RQ$ slope increases as $\epsilon_\text{min}$ decreases, shrinking the range of $\dot{\nu}$. The curve $PQ$ shifts to the left as $\epsilon_\text{max}$ increases, increasing $\lvert \dot{\nu} \rvert$, and hence lowering $\nu$.

Let us now see what happens when we vary the minimum and maximum magnetic field, $B_\text{min}$ and $B_\text{max}$. The absolute value of the $RS$ slope decreases as $B_\text{min}$ increases, stretching the box sideways as we retreat from the gravitational-wave dominated limit. The blue box is always bounded above by the $\langle n \rangle = 5$ age line. It shrinks, and flattens as the role of $Q_1$ diminishes.

We now discuss the purple and red boxes in which $Q_2$ dominates. The region bounded by the purple box has $B = 10^{13}$\,G, and $n = 2.3$, whereas the red box has the same $B$, but $n = 3$. Reducing $n$ increases the spin-down rate by a factor of ($\pi R_{\star}/c)^{n-3}$. Hence, for the same $\epsilon$ and $B$, the purple box covers a smaller range of $\nu$ than the red box. Both are considerably smaller than the blue box for the same range of $\epsilon$ and $\nu_b$. 
Again, if $\epsilon_\text{max}$ increases, the purple and red boxes expand downwards.
In Figure \ref{ffdot_large}, we choose to plot the purple box with $n = 2.3$ because it lies partially within the sensitivity range of the search. Importantly, $\nu$ and $\dot{\nu}$ end up outside the search sensitivity range for $n < 2.3$ or $B > 10^{13}$\,G, restricting the range of astrophysical birth scenarios that our search is sensitive to.

\begin{figure*}
\centering
\scalebox{0.7}{\includegraphics{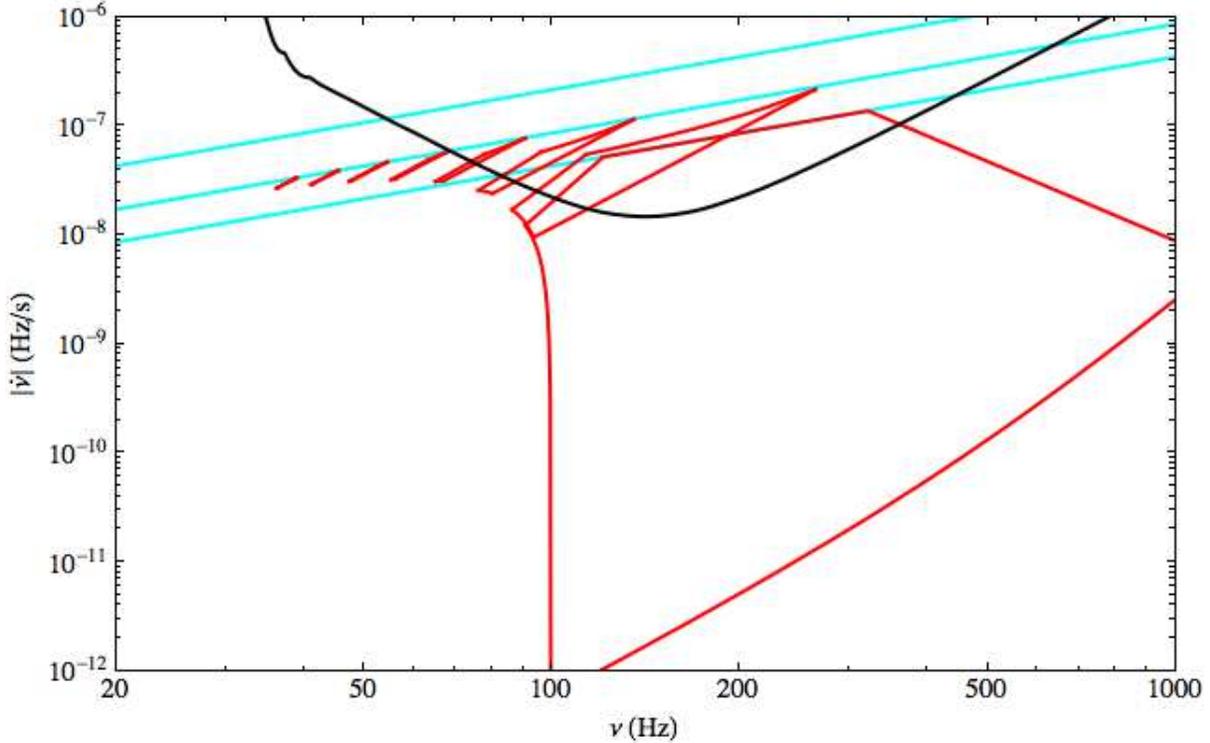}}
\caption[Effect of varying $B$ on the $\nu, \dot{\nu}$ plane. ]{Final states ($\nu$, $\dot{\nu}$) calculated from equation (\ref{eq:numodel}) on the $\nu$-$\lvert \dot{\nu} \rvert$ plane, for a range of magnetic field strengths. The eight red boxes surround regions which have $n = 3$ and cover the same range of $\epsilon$ and $\nu_b$ as Figure \ref{ffdot_large}. Their magnetic fields range from $B = 10^{11}$G (largest box) to $B = 10^{14.5}$ G (smallest box). }
\label{ffdot-brange-large}
\end{figure*}

 The range of $\nu$ covered in the $Q_2$-dominated limit is sensitive to $B$. 
In Figure \ref{ffdot-brange-large}, we show explicitly how varying $B$ affects $\nu,\dot{\nu}$. We plot eight red boxes, for $10^{11}$\,G (largest box) $\leq B \leq 10^{14.5}$\,G (smallest box), and $n = 3$.  
As $B$ increases, the red boxes shift to the left. For $B \geq 5 \times 10^{13}\rm{G}$, the box falls out of the sensitivity range of the search. Also, the boxes shrink as $B_\text{max}$ increases. This happens because as $B$ increases, $\dot{\nu}_\text{EM}$ increases. For $B \gtrsim 10^{14}$\,G, we find $\nu \ll \nu_b$ after 19 years, and the boxes end up on the $\langle n \rangle = 3$ line.  
All the red boxes are bounded above by the $\langle n \rangle = 3$ age line. 

Figures \ref{ffdot_large} and \ref{ffdot-brange-large} provide constraints on the detectable range of $\epsilon, B, n,$ over a broad range of $\nu_b$. We conclude that, in preparing to select the search templates, it is sensible to consider the parameter range $10^{-5} \geq \epsilon$, $B \leq 10^{13}$\,G, $2.3 \leq n \leq 5$. A more detailed breakdown of the detectable and computationally feasible parameter ranges is presented in Section \ref{sec:templates}.
Note that even though the particular boxes drawn as examples in Figures \ref{ffdot_large} and \ref{ffdot-brange-large} do not cover the entire region between the sensitivity curve and the $\langle n \rangle = 1.8$ line, one can potentially reach any point in that region with some combination of $n$, $\epsilon$ and $B$. Also, each $(\nu, \dot{\nu})$ pair in the figures can be reached by an infinite set of combinations ($\epsilon$, $B$, $n$ and $\nu_b$). However, there are combinations of $\nu$ and $\dot{\nu}$ which are allowed in principle by age-based indirect limits but which cannot be reached from $\nu_b$ with realistic choices of $\epsilon, B$, and $n$.

\section{Template spacing}
\label{metric}

The cross-correlation search for SNR 1987A is computationally limited rather than sensitivity limited over much of the parameter space in Figures \ref{ffdot_large} and \ref{ffdot-brange-large}. Therefore, the placement of templates is crucial. If the template grid is too coarse, the risk of missing the signal increases; if it is too fine, time is wasted searching redundant templates. In order to compute the optimal spacing, we construct a phase metric \citep{bala96, owen96} which computes the signal-to-noise ratio as a function of template spacing along each axis of the four-dimensional parameter space ($\nu, \epsilon, B, n$). The coherent phase metric for the conventional Taylor-expansion phase model is widely used in LIGO in both coherent and semi-coherent searches \citep{brady00, prix07, wette08}, although its semi-coherent form has not been fully investigated. In this section, we derive the semi-coherent phase metric for the astrophysical phase model defined by integrating (\ref{eq:numodel}). We also estimate the range of detectable spin-down values as well as magnetic field, ellipticity and braking index values given a computationally feasible number of templates. 

\subsection{Semi-coherent phase metric}
\label{sec:phasemetric}
When searching a template grid, it is extremely unlikely that one particular set of parameters will match the true signal exactly. What we have instead is a set of guessed parameters $\boldsymbol{\theta} + \Delta \boldsymbol{\theta}$, describing the closest match, which are offset from the true values by a small amount, $\Delta \boldsymbol{\theta}$. 
For a given set of guessed parameters, the power spectrum of a time-coincident SFT pair is 
\begin{equation}
\label{eq:sftmismatch} \mathcal{P}(\boldsymbol{\theta}, \Delta \theta) =  \frac{2\mathcal{A}}{\sqrt{\Delta T}} \left \lvert \int_{\Tstart}^{\Tstart+ \Delta T}\,dt\, e^{i \Delta \Phi(t)} \right\rvert^2 ,
\end{equation}
where $\Delta \Phi(t) =  \Phi(t, \boldsymbol{\theta} + \Delta \boldsymbol{\theta}) - \Phi(t, \boldsymbol{\theta})$ is the mismatch between the actual and guessed phases, $\Tstart$ is the time at the midpoint of the SFT, and $\mathcal{A}$ is the gravitational wave amplitude. 

The mismatch between (\ref{eq:sftmismatch}) and the power spectrum of the SFT pair if $\Delta \boldsymbol{\theta} = 0$ is defined to be
\begin{equation}
m(\boldsymbol{\theta}, \Delta \boldsymbol{\theta}) = 1 - \frac{\mathcal{P}(\boldsymbol{\theta}, \Delta \boldsymbol{\theta})}{\mathcal{P}(\boldsymbol{\theta},0)}
\end{equation}
and is related to the semi-coherent phase metric $s_{ij}$ by
\begin{equation}
\label{eqn:mismatch}
m(\boldsymbol{\theta}, \Delta \boldsymbol{\theta}) =  s_{ij} (\boldsymbol{\theta}) \Delta \theta^i \Delta \theta^j,
\end{equation}
where $1 \leq i,j \leq 4$ label the various search parameters. 

For the cross-correlation search, we have $\boldsymbol{\theta} = \left(\nu, Q_1, Q_2, n\right)$. Hence, for a given mismatch $m$, the minimum (i.e. most conservative) template spacings are given by $\Delta \nu (\boldsymbol{\theta}) = \sqrt{m/s_{00} (\boldsymbol{\theta})}, \Delta Q_1 (\boldsymbol{\theta}) = \sqrt{m/s_{11} (\boldsymbol{\theta})}, \Delta Q_2 (\boldsymbol{\theta}) = \sqrt{m/s_{22} (\boldsymbol{\theta})}, \Delta n (\boldsymbol{\theta}) = \sqrt{m/s_{33} (\boldsymbol{\theta})}$. Note that it may be possible to do better (i.e. expand the spacing) by taking advantage of the covariances between parameters embodied in the metric through (\ref{eqn:mismatch}); this issue deserves further study.

In order to calculate $s_{ij}$, we must first calculate the \textit{coherent} phase metric, defined to be 
\begin{equation}
\label{eq:gij}
g_{ij} = \langle \partial_i \Delta \Phi \partial_j \Delta \Phi \rangle - \langle \partial_i \Delta \Phi \rangle \langle \partial_j \Delta \Phi \rangle,
\end{equation}
with $\langle ... \rangle = \frac{1}{\Tlag}\int_{\Tstart}^{\Tstart + \Tlag}\,dt\,...$ and $\partial_i \Delta \Phi = \partial \Delta \Phi/\partial \Delta \theta^i$ evaluated at $\Delta \boldsymbol{\theta} = 0$. Calculating $g_{ij}$ analytically by integrating (\ref{eq:numodel}) is non-trivial. However, a good approximation results if we integrate (\ref{eq:numodel}) \textit{separately} for the gravitational-wave and electromagnetic torques, and combine the answers in quadrature. Details of the calculation are shown in Appendix A. In brief, tracking the gravitational-wave and electromagnetic spin down separately yields two ``sub-metrics'', one comprising $\nu$ and $Q_1$ (gravitational) and the other comprising $\nu, Q_2,$ and $n$ (electromagnetic). Diagonal elements of $s_{ij}$ can be obtained by summing the two sub-metrics.

The semi-coherent metric $s_{ij}$ is the average of the coherent metric from $\Tstart = 0$ to $\Tstart = \Tobs$, where $\Tobs$ is the entire observation time spanned by all SFT pairs. It is defined to be $s_{ij} = \left(\Tobs\right)^{-1} \int_{0}^{\Tobs} d\Tstart\, g_{ij}$. From Appendix B, the diagonal elements of the semi-coherent metric are:
\begin{eqnarray}
\label{eqn:s00}
\nonumber s_{00} &\approx& 10 \Tlag^2 \Tobs^2 \left( \frac{5}{72} K_1^2 \nu^8 Q_1^2 - \frac{1}{36} K_1 K_2 n \nu^{n+3} Q_1 Q_2  \right.\\
&&\left. + \frac{1}{360} K_2^2 n^2 \nu^{2n-2} Q_2^2\right) ,\\
s_{11} &\approx& \frac{K_1^2 \nu^{10}}{36} \Tlag^2 T^2_\text{obs},\\
s_{22} &\approx& \frac{K_2^2 \nu^{2n}}{36} \Tlag^2 T^2_\text{obs},\\
\label{eqn:s33} s_{33} &\approx& \frac{K_2^2 \log \left(\nu\right)^2 \nu^{2n} Q_2^2}{36} \Tlag^2 T^2_\text{obs},
\end{eqnarray}
with $K_1 = K_1 (\nu, Q_1)$ and $K_2 = K_2 (\nu, Q_2, n)$. For pure gravitational-wave and electromagnetic spin down, we have $(K_1, K_2) = (1, 0)$ and $(0, 1)$ respectively. The full expressions for (\ref{eqn:s00})--(\ref{eqn:s33}) are presented in Appendix B. 
Note that in (\ref{eqn:s00})--(\ref{eqn:s33}), all frequency terms are normalised by $\nu_\text{ref}$. For clarity, we have set $\nu_\text{ref} = 1$\,Hz and do not display it.  

In Appendix C, we estimate the phase error which accumulates after a time $\Tlag$ from mismatches in $\nu, Q_1, Q_2$, and $n$. We find that it scales with $\Tlag$ similarly to (\ref{eqn:s00})--(\ref{eqn:s33}) for $Q_1$ and $Q_2$. For $\nu$, the phase error scales instead as $\Tlag$, and for $n$, it scales as $\Tlag \log\left[1 + (n-1)Q_2 \Tlag \nu^{n-1}\right]$. In a semi-coherent search, the phase needs to be tracked to within $\pi/4$ over the interval $\Tlag$, not $\Tobs$, unlike in fully coherent searches. Across the entire observation time $\Tobs$, we require only that the frequency of the signal be tracked to within $1/\Delta T$. This adds an overall $\Tobs^2$ dependence to (\ref{eqn:s00})--(\ref{eqn:s33}). 

\subsection{Computational cost of the search}
\label{sec:templates}
 The run-time of the search code is proportional to $N_\text{pairs} N_\text{total}$, where $N_\text{total}$ is the total number of templates required to search the parameter space. Trials with $N_\text{pairs} = 10^5$ comprising 1 year's worth of SFTs (from the two interferometers H1 and L1), and $\Tlag = 1$\,hour take $\sim 1$\,s per template on a single, 1-gigaflop computational node. We can therefore search $\sim 10^{9}$ templates in a realistic run using $10^3$ nodes over two weeks.

We now compare the computational cost of the astrophysical phase model (\ref{eq:numodel}) against the Taylor-expansion model (\ref{eq:phitaylor}). The semi-coherent metric for the latter model is not well studied, however recent work has yielded analytic expressions for the metric \citep{pletsch09,pletsch10}. Based on these expressions, we can estimate the number of templates in the following way.

Firstly, we consider the number of templates required to track the phase coherently over a time $\Tlag$. 
For the $k$-th frequency derivative in the Taylor expansion model, the corresponding diagonal term of the \textit{coherent} metric scales as ($g_{ij}$)$_\text{coh}^{(k)} \propto \Tlag^{2k + 2}$ \citep{whitbeck06}. The number of templates required to track the $k$-th frequency derivative coherently is then $N_{k} \propto \sqrt{(g_{ij})^{(k)}_\text{coh}} \propto \Tlag^{k+1}$.
 The total number of templates required for each coherent chunk of length $\Tlag$ is therefore given by $N_\text{coh}  = \prod^s_{k=0} N_k$, i.e. $N_\text{coh} \propto \prod^s_{k=0} \Tlag^{k+1}$, where $s$ is the number of frequency derivatives required to track the gravitational wave phase (see Section \ref{searchparams}). Now, assume that over a time $\Tobs$, we sum a number of chunks incoherently, approximately proportional to $N_\text{chunks} \propto \Tobs/\Tlag$.\footnote{We emphasize that this is only an approximate estimate, as the cross-correlation method sums SFT pairs separated by a time up to and including $\Tlag$. Strictly speaking, $N_\text{chunks} > \Tobs/\Tlag$.} Now, using the semi-coherent metric \citep{pletsch09, pletsch10}, the number of templates required for $s$ frequency derivatives is proportional to $\gamma_s N_\text{coh}$, where $\gamma_s$ is a `refinement factor' which scales as $N_\text{chunks}^{s(s+1)/2}$.
The total number of templates is then approximately
\begin{eqnarray}
N_\text{total} &\propto&  N_\text{chunks}^{s(s+1)/2} \prod^s_{k=0} \Tlag^{k+1}\\
&\propto& \left(\frac{\Tobs}{\Tlag}\right)^{s(s+1)/2}  \prod^s_{k=0} \Tlag^{k+1}.
\end{eqnarray}
For the range of $(\nu_b, \epsilon, B, n)$ considered in Section \ref{sec:historical}, for $\Tlag = 1$\,hr, we must track terms up to and including $\nu^{(4)}$ in (\ref{eq:phitaylor}) in order to keep the phase error overall below $\pi/4$. This gives $N_\text{total} \propto \Tobs^{10}\Tlag^5$.

Under the astrophysical phase model, we estimate $N_\text{total} = N_\nu N_{Q1} N_{Q2} N_n$ from (\ref{eqn:s00})--(\ref{eqn:s33}), where the subscripts denote the number of templates required for each individual parameter, e.g. $N_\nu \approx \nu/\Delta \nu$. As (\ref{eqn:s00}) yields different results for $N_\nu$ in the gravitational and electromagnetic limits, we bound $N_\nu$ by taking it to be the sum of squares of the two limits, i.e. $N_\nu = \left[N_{\nu, (K_1, K_2) = (1,0)}^2 + N_{\nu, (K_1,K_2) = (0,1)}^2\right]^{1/2}$. For a given mismatch $m$, we obtain
\begin{equation}
\label{templateestimate} N_{\text{total}} \propto m^{-2} n \nu^{10 + 3n} \log(\nu) Q_1 Q_2^2 \left[Q_1^2 + n^2 Q_2^2\right]^{1/2} \Tlag^4 \Tobs^4
\end{equation}

Equation (\ref{templateestimate}) is an approximate result, achieved by combining the two sub-metrics used in equations (\ref{eqn:s00})--(\ref{eqn:s33}). It should be regarded as a rule of thumb. If gravitational-wave spin down dominates, we have $s_{22} = s_{33} = 0$, $\tau_c = (4 Q_1 \nu^4)^{-1}$, and hence
\begin{equation}
N_{\text{total}} \propto m^{-1} \nu^{2} \tau_c^{-2} \Tlag^2 \Tobs^2.
\end{equation}
If electromagnetic spin down dominates, we have $s_{11} = 0$, $\tau_c = \left[\left(n-1\right)Q_2 \nu^{n-1}\right]^{-1}$, and hence
\begin{equation}
N_{\text{total}} \propto m^{-3/2} n^2 \nu^{3} \log(\nu) \left[(n-1) \tau_c\right]^{-3} \Tlag^3 \Tobs^3.
\end{equation}

The required template spacing therefore varies dramatically across the astrophysical parameter range. To illustrate, let us consider $0.1\,\text{kHz} \leq \nu \leq 1$\,kHz, $10^{-22}\,\rm{s}^3 \leq Q_1 \leq 10^{-18}\,\rm{s}^3$, $10^{-21}\,\rm{s}^2 \leq Q_2 \leq 10^{-13}\,\rm{s}^2$, and $2.3 \leq n \leq 3.0$, and hence  $8 \times 10^{-6} \leq \epsilon \leq 8 \times 10^{-4}$, $4 \times 10^9\,\text{G} \leq B \leq 4 \times 10^{13}$\,G. We assume a mismatch $m$ of 0.2. The required resolutions in the four search parameters range across
\begin{eqnarray}
\label{eq:deltanu}
 2.935 \times 10^{-4} \leq &\Delta \nu/\rm{Hz}& \leq 9.632 \times 10^{-4},\\
2.973 \times 10^{-26} \leq &\Delta Q_1/\rm{s}^3& \leq 3.685 \times 10^{-22},\\
3.448 \times 10^{-20} \leq &\Delta Q_2/\rm{s}^2& \leq 2.120 \times 10^{-16},\\
\label{eq:deltan} 8.674 \times 10^{-8} \leq &\Delta n& \leq 1.166 \times 10^4
\end{eqnarray}
in this search volume. The number of templates required for each parameter is its range divided by its bin resolution. If the bin resolution is larger than its range, we require only one template. Equations (\ref{eq:deltanu})--(\ref{eq:deltan}) imply a total number of templates between $2.958 \times 10^5 \leq N_\text{total} \leq 4.347 \times 10^{26}$ to cover the entire parameter space. Smaller values of $\nu, Q_1, Q_2$ and $n$ require fewer templates to cover their neighbourhood.

Unfortunately, given the computational restrictions that we face, we cannot search the entire region of astrophysical parameters in Figure \ref{ffdot_large}. In the following analysis, we therefore divide each axis in parameter space into (say) ten bins, i.e. a $10 \times 10 \times 10 \times 10$ hypercubic grid containing $10^4$ ``boxes", and calculate the localised resolution at the centroid of each box. The grid is spaced logarithmically along $\epsilon$ and $B$ to cover $Q_1$ and $Q_2$ in a representative fashion. Only those boxes requiring $N \lesssim 10^9$ are practical to search.

\subsection{Astrophysical upper limits}
\label{sec:upperlimits}
In this section, we combine the estimates of sensitivity and computational cost in Sections \ref{sec:ccsensitivity} and \ref{sec:templates} respectively to identify the ranges of the astrophysical parameters $B$ and $\epsilon$ that can be probed by a realistic search.
In the event of a non-detection, upper limits on $B$ and $\epsilon$ can be placed.

\begin{figure*}
\centering
\scalebox{0.56}{\includegraphics{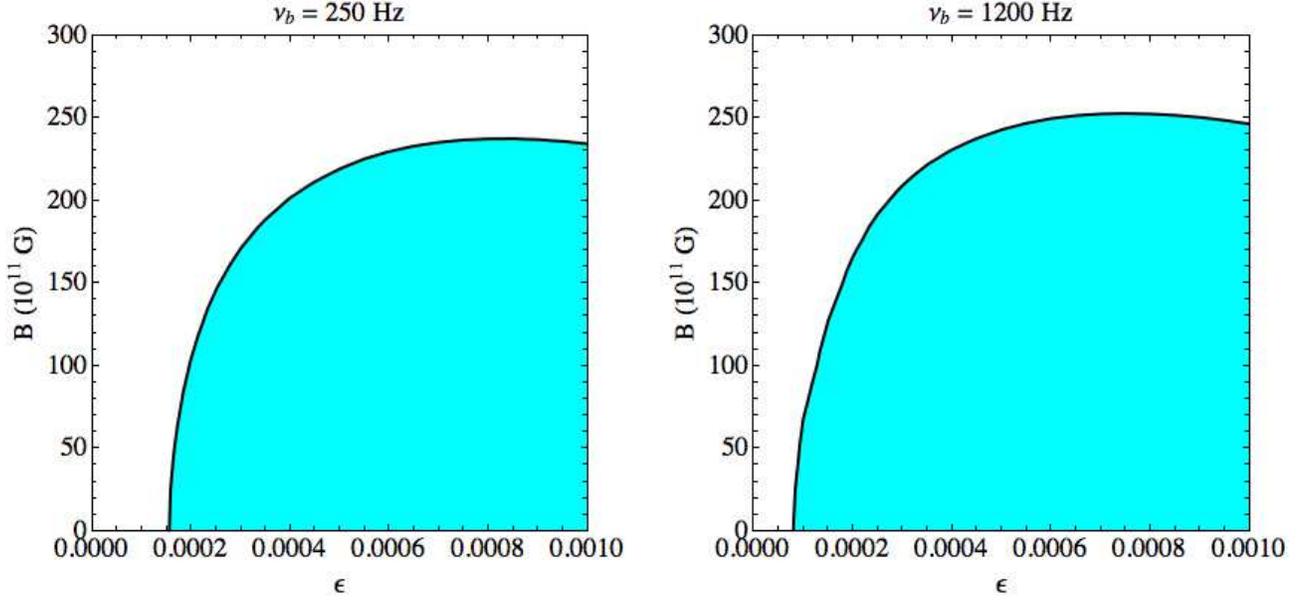}}
\caption[Contour plots of $h_0$ as a function of $\epsilon$ and $B (10^{11}\,\rm{G})$ for SNR 1987A for birth frequencies of $\nu_b = $ 250 and 1200\,Hz.]{Contour plots of $h_0$ as a function of $\epsilon$ and $B (10^{11}$\,G) for SNR 1987A for values of $\nu_b = $ 250 (left panel) and 1200\,Hz (right panel). We assume $n = 3$ and a pulsar age of 19 years, as the S5 run began in 2006. The cyan shaded areas correspond to $h_0 \geq h_\text{th}$, where $h_\text{th}$ is defined in (\ref{eq:hth}).}
\label{q1vsq2}
\end{figure*}

 We solve (\ref{eq:numodel}) for a range of $\nu_b, \epsilon,$ and $B$, and calculate the characteristic wave strain $h_0$ from (\ref{hstrain}). Figure \ref{q1vsq2} displays contours of $h_0$ versus $B$ and $\epsilon$ for $n=3$ at two frequencies corresponding to $\nu_b = 300$\,Hz and $\nu_b = 1200$\,Hz.
The cyan shaded areas indicate where $h_0 \geq h_\text{th}$. The search is sensitive to a larger range of $\epsilon$ and $B$ as $\nu_b$ rises. This occurs because the search sensitivity peaks at $\nu \approx 150$\,Hz. For small $\nu_b$ and large $\epsilon$ and $B$, the pulsar spins down after $\tau_c = $ 19\,yr to give $\nu < 150$\,Hz. In the best case scenario, for $\nu_b = 1200$\,Hz, upper and lower limits on the magnetic field and ellipticity of $B \lesssim 2.5 \times 10^{13}$\,G and $\epsilon \gtrsim 8 \times 10^{-5}$ can be achieved. 

\begin{figure*}
\centering
\scalebox{0.6}{\includegraphics{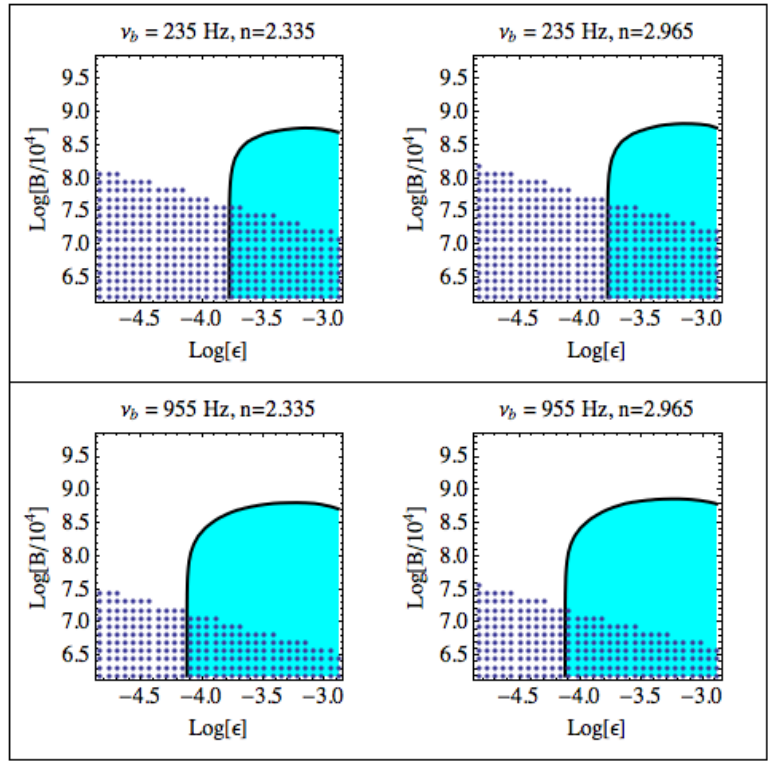}}
\caption[Log-log contour plots of $h_0$ as a function of $\epsilon$ and $B$ for a range of birth frequencies, $\nu_b$, and braking indices $n$.]{Log-log contour plots of $h_0$ as a function of $\epsilon$ and $B$ ($10^4\,\rm{G})$ for birth frequencies $\nu_b = 235$\,Hz (top panel) and $\nu_b = 955$\,Hz (bottom panel), and a range braking indices, $n$. The frequency of the signal $\nu$ is obtained by solving (\ref{eq:numodel}), and integrating over $\tau_c =$ 19\,yr. The cyan shaded areas indicate the regions in which $h_0 \geq h_\text{th}$, where $h_\text{th}$ is defined in (\ref{eq:hth}). The panels are arranged in pairs. Each pair shows $n = 2.335$ (left) and $n = 2.965$ (right). The dark blue dots indicate parameter combinations for which one has $N \leq 10^{9}$.}
\label{ntempslog}
\end{figure*}

Unfortunately, the number of search templates required to cover the shaded region in Figure \ref{q1vsq2} is prohibitively large, as discussed in Section \ref{sec:templates}.
Figure \ref{ntempslog} shows both sensitivity and computational cost. Regions in which the search is sensitive (i.e. $h_0 \geq h_\text{th}$) for given $\nu_b$ and $n$ are shaded in cyan. Overplotted as dark blue dots are the central coordinates of our grid boxes with $N \leq 10^{9}$. The panels correspond to a range of \textit{birth} frequencies, $\nu_b$, and are grouped in pairs: $n = 2.335$ (left panel in pair) and $n = 2.965$ (right panel in pair). The top pair shows the sensitivity and computational cost for $\nu_b = 235$\,Hz, whereas the bottom pair corresponds to $\nu_b = 955$\,Hz (bottom right). 

Figure \ref{ntempslog} shows that the search sensitivity increases with $\nu_b$ and $\epsilon$, and decreases with $B$. On the other hand, the computational efficiency of the search decreases with $\nu_b$ and $\epsilon$, and increases with $B$. Even so, there is substantial overlap between the regions in which the search is sensitive and the regions which are computationally permissible. We note that as each individual dot in Figure \ref{ntempslog} represents a region in which $N \leq 10^{9}$, it is not feasible to search over \textit{all} the dotted areas, as this would mean $N_\text{total} \geq 10^{9}$. Therefore, when implementing the search, we will choose an appropriate range of parameters such that $N_\text{total} \lesssim 10^{9}$, using Figure \ref{ntempslog} as a guide. 

\begin{table}
\centering
\begin{tabular}{ccc}
\hline
\hline
$\nu_b$\,(kHz) & $\epsilon$ & $B$\,($10^{11}$\,G)\\
\hline
0.19--0.28 &  $\gtrsim 1.6 \times 10^{-4}$ & $\lesssim$ 2.0\\
0.28--0.55 &  $\gtrsim 1.0 \times 10^{-4}$ & $\lesssim 1.3$\\
0.55--1.00&  $\gtrsim 7.9 \times 10^{-5}$ & $\lesssim 0.8$\\
\hline
\end{tabular}
\caption[Table of $\nu_b$, $\epsilon$, and $B$ which are detectable by the cross-correlation search.]{Table of $\nu_b$, $\epsilon$, and $B$ ranges (approximate) which are detectable by the cross-correlation search for SNR 1987A using LIGO S5 data, for $2.3 \leq n \leq 3.0$. The numbers in the table are based on the regions in which the computationally feasible (dark blue dots) and search-sensitive (cyan shaded) regions overlap in Figure \ref{ntempslog}. We assume standard values for the neutron star mass and radius, i.e. $M_\star = 1.4 M_\odot, R_\star = 10$\,km, and $n = 3$.} 
\label{tab:results}
\end{table}

Table \ref{tab:results} summarises the approximate range of $\nu_b$, $\epsilon$, and $B$ in which the two regions in Figure \ref{ntempslog} overlap. If the pulsar in SNR 1987A was born with a frequency between 0.19\,kHz and 0.28\,kHz, the search is sensitive to $\epsilon \gtrsim 1.6 \times 10^{-4}$ and $B \lesssim 2.0 \times 10^{11}$\,G. This range narrows as $\nu_b$ increases; for birth frequencies between 0.55\,kHz and 1.00\,kHz, the search is sensitive to $\epsilon \gtrsim 7.9 \times 10^{-5}$ and $B \lesssim 0.8 \times 10^{11}$\,G. We note that these estimates, derived from the limits on $Q_1$ and $Q_2$, assume the standard values for the neutron star mass and radius, $M_\star = 1.4 M_\odot$ and $R_\star = 10$\,km. It is possible that SNR 1987A contains a low-mass neutron star with $M_\star \approx 0.13 M_\odot$ \citep{imshenik92}, in which case the limit on the ellipticity, for $\nu_b = 1.00$\,kHz, would be $\epsilon \gtrsim 2.5 \times 10^{-4}$.

We now comment briefly on the relevance of these limits. The range of $B$ listed in Table \ref{tab:results} is within the expected theoretical range discussed in Section \ref{sn1987Adetails} \citep{michel94, ogelman04}. The range of $\epsilon$ listed in Table \ref{tab:results}, however, is larger than the maximum ellipticity sustainable by the unmagnetized neutron star crust for many equations of state. For example, conventional neutron stars are expected to support $\epsilon \leq 10^{-6}$, while hybrid quark-baryon or meson-condensate stars can support $\epsilon \leq 10^{-5}$ \citep{ushomirsky00, owen05, horowitz09}. However, some exotic models do allow for larger ellipticities. Solid strange quark stars are predicted to be able to sustain $\epsilon \leq 6 \times 10^{-4}$ \citep{owen05}. For low-mass neutron stars, the limit is $\epsilon \leq 5 \times 10^{-3}$ \citep{imshenik92, horowitz10}. We note also that these limits apply only to elastically supported deformations; magnetically supported deformations can be larger \citep{melatos07, akgun08, haskell08}. Therefore, even placing the relatively large upper limit of $\epsilon \lesssim 10^{-4}$ on the putative neutron star in SNR 1987A will be useful to some degree in constraining its mass and/or equation of state.

\section{Conclusion}
\label{sec:ccconclusion}
In this paper, we describe the steps taken to quantify the astrophysical significance of a cross-correlation search for the supernova remnant SNR 1987A in LIGO S5 data.
\begin{itemize}
\item We estimate the theoretical sensitivity of the cross-correlation search, and compare it to the conservative, age-based, wave strain estimate. In the frequency band 75\,Hz $\lesssim \nu \lesssim$ 450\,Hz, the age-based estimate lies above the detection threshold.
\item  We introduce an alternative to the Taylor expansion model of the gravitational wave phase based on a set of four astrophysical search parameters ($\nu, \epsilon, B, n$). The new phase model renders a search for a neutron star like SNR 1987A with a high spin-down rate computationally feasible.
\item To estimate the optimal template spacing for the search, we calculate the semi-coherent phase metric corresponding to this astrophysical model. 
\item We place detection limits on $\epsilon$ and $B$ for a range of birth spin frequencies, 0.1\,kHz $\leq \nu_b \leq$ 1.2\,kHz.
\end{itemize}

With the required template spacing and current computational capabilities discussed in Section \ref{sec:templates}, we will be able to search up to approximately $10^{9}$ templates. In the event of a non-detection, considering the parameter range discussed in this paper and assuming the standard neutron star mass and radius, we expect to place the following limits on the pulsar's ellipticity and magnetic field: $\epsilon \leq 8 \times 10^{-5}$, $B \geq 2.0 \times 10^{11}$\,G. The search is also expected to be sensitive to electromagnetic braking indices $2.3 \leq n \leq 3.0$. Its greatest weakness remains that it assumes $n$ to be constant throughout the semi-coherent integration. Constant $n$ is the simplest possible astrophysical scenario, and it certainly deserves to be considered in its own right, in view of the overwhelming computational cost of a variable-$n$ search. Nevertheless, it is vital to recognize that the constant-$n$ hypothesis covers a small fraction of the astrophysical parameter space.

A search using gravitational wave data is anticipated to begin soon and would be the first application of the cross-correlation method to a continuous wave search.

\section*{Acknowledgements}
CC acknowledges the support of an Australian Postgraduate Award and the Albert Shimmins Memorial Fund. JTW acknowledges the support of NSF grant PHY-0855494, the College of Science at Rochester Institute of Technology, and the German Aerospace Center (DLR). This paper has been designated LIGO Document No. LIGO-P1000089-v3.

\begin{onecolumn}
\section*{Appendix A: Calculation of the coherent metric $g_{ij}$}
This appendix details the calculation of the diagonal terms of the coherent metric, $g_{ij}$ (\ref{eq:gij}). We start by evaluating the frequency $\nu(t)$ at time $t$, by assuming that $\nu(t)$ is a simple sum of separate, independent contributions from gravitational-wave and electromagnetic spin down:
\begin{eqnarray}
\label{eq:nuoft} \nu(t) &=& K_1 \int -Q_1 \nu(t)^5 dt + K_2 \int -Q_2 \nu(t)^n dt\\
\label{eq:nuoft2} &=&\frac{K_1 \nu}{\left(1 + 4 Q_1 \nu^4 t\right)^{1/4}} + \frac{K_2 \nu}{\left[1 + (n-1) Q_2 \nu^{n-1} t\right]^{1/n-1}}.
\end{eqnarray}
Here, $K_1$ and $K_2$ are constants which satisfy $K_1 + K_2 = 1$, and the search parameters $\nu, Q_1, Q_2,$ and $n$ are defined at a reference time $t_0$. Recall that $\nu$ is normalised by $\nu_\text{ref}$, which we set to 1\,Hz and do not write down, for simplicity. The first term in (\ref{eq:nuoft2}) follows from the first integral in (\ref{eq:nuoft}) by assuming $Q_2 = 0$. The second term in (\ref{eq:nuoft2}) follows from the second integral in (\ref{eq:nuoft}) by assuming $Q_1 = 0$. Needless to say, the exact solution for $\nu(t)$ follows from solving (\ref{eq:numodel}) self-consistently for $Q_1 \neq 0, Q_2 \neq 0$, but this is too difficult to solve analytically. As the phase metric calculation is useful only in an analytic form, we adopt the approximation in (\ref{eq:nuoft}). 

The phase at time $t$ is given by,
\begin{eqnarray}
\Phi(t, \boldsymbol{\theta}) &=& \int^{t + t_0}_{t_0} dt\, \nu(t)\\
\nonumber  &=& \frac{K_1 \left[1 + 4 Q_1 \nu^4 \left(t + \frac{\mathbf{r}.\mathbf{n}}{c}\right)\right]^{3/4} }{3 Q_1 \nu^3}  - \frac{K_1 \left(1 + 4 Q_1 \nu^4 t_0\right)^{3/4}}{3 Q_1 \nu^3}\\
 && + \frac{K_2 \left[1 + (n-1) Q_2 \nu^{n-1} \left(t + \frac{\mathbf{r}.\mathbf{n}}{c}\right)\right]^{\frac{2-n}{1-n}}}{(n-2) Q_2 \nu^{n-2}}  - \frac{ K_2 \left[1 + (n-1) Q_2 \nu^{n-1} t_0\right]^{\frac{2-n}{1-n}}}{(n-2) Q_2 \nu^{n-2}}.
\end{eqnarray}

We can expand each term in the regimes $(Q_1 \nu^4)^{-1} \gg t$ and $(Q_2 \nu^{n-1})^{-1} \gg t$, giving
\begin{eqnarray}
\label{eq:phitheta}
\nonumber \Phi(t, \boldsymbol{\theta}) &=& K_1 \nu \left(t + \frac{\mathbf{r}.\mathbf{n}}{c} - t_0\right) - \frac{K_1}{2}Q_1 \nu^5 \left[\left(t + \frac{\mathbf{r}.\mathbf{n}}{c}\right)^2 - t_0^2\right]  \\
&& + K_2 \nu \left(t + \frac{\mathbf{r}.\mathbf{n}}{c} - t_0\right) - \frac{K_2}{2} Q_2 \nu^{n} [(t + \frac{\mathbf{r}.\mathbf{n}}{c})^2 - t_0^2],
\end{eqnarray}
and
\begin{eqnarray}
\label{eq:phideltatheta}
\nonumber \Phi(t, \boldsymbol{\theta} + \Delta \boldsymbol{\theta}) &=& K_1 (\nu + \Delta \nu) \left(t + \frac{\mathbf{r}.\mathbf{n}}{c} - t_0\right)  - \frac{K_1}{2}(Q_1 + \Delta Q_1) (\nu + \Delta \nu)^5 \left[\left(t + \frac{\mathbf{r}.\mathbf{n}}{c}\right)^2 - t_0^2\right]  \\
&& + K_2 (\nu + \Delta \nu)\left(t + \frac{\mathbf{r}.\mathbf{n}}{c} - t_0\right) - \frac{K_2}{2} (Q_2 + \Delta Q_2) (\nu + \Delta \nu)^{n + \Delta n} \left[\left(t + \frac{\mathbf{r}.\mathbf{n}}{c}\right)^2 - t_0^2\right].
\end{eqnarray}
Subtracting (\ref{eq:phitheta}) from (\ref{eq:phideltatheta}) gives 
\begin{eqnarray}
\label{eq:deltaphi}
\nonumber \Delta \Phi (t) &=& \Delta \nu (K_2 + K_1) \left(t + \frac{\mathbf{r}.\mathbf{n}}{c} - t_0\right)  \\
\nonumber && - \frac{K_1}{2} \left[\left(t+\frac{\mathbf{r}.\mathbf{n}}{c}\right)^2 - t_0^2\right] \left[(\nu + \Delta \nu)^5 (Q_1 + \Delta Q_1) - \nu^5 Q_1\right]  \\
 && - \frac{K_2}{2} \left[\left(t+\frac{\mathbf{r}.\mathbf{n}}{c}\right)^2 - t_0^2\right]\left[ (\nu + \Delta \nu)^{n+\Delta n} (Q_2 + \Delta Q_2)  - \nu^n Q_2\right]
\end{eqnarray}

We now take the derivative of (\ref{eq:deltaphi}) with respect to $\Delta \nu$, $\Delta Q_1$, $\Delta Q_2$, and $\Delta n$.  We have
\begin{eqnarray}
\label{eq:dphidnu} \partial_{\Delta \nu} \Delta \Phi(t)\rvert_{\Delta \theta = 0} &=& (K_1 + K_2) \left(t + \frac{\mathbf{r}.\mathbf{n}}{c} - t_0\right)  - \frac{5}{2} \nu^4 Q_1 \left[\left(t + \frac{\mathbf{r}.\mathbf{n}}{c}\right)^2 - t_0^2\right]   - \frac{K_2}{2} n \nu^{n-1} Q_2 \left[\left(t - \frac{\mathbf{r}.\mathbf{n}}{c}\right)^2 - t_0^2\right] 
\end{eqnarray}
\begin{eqnarray}
\partial_{\Delta Q_1} \Delta \Phi(t)\rvert_{\Delta \theta = 0} &=& -\frac{K_1}{2} \nu^5 \left[\left(t + \frac{\mathbf{r}.\mathbf{n}}{c}\right)^2 - t_0^2\right]
\end{eqnarray}
\begin{eqnarray}
\partial_{\Delta Q_2} \Delta \Phi(t)\rvert_{\Delta \theta = 0} &=& -\frac{K_2}{2} \nu^n \left[\left(t + \frac{\mathbf{r}.\mathbf{n}}{c}\right)^2 - t_0^2\right]  
\end{eqnarray}
\begin{eqnarray}
\label{eq:dphidn} \partial_{\Delta n} \Delta \Phi(t)\rvert_{\Delta \theta = 0} &=& -\frac{K_2}{2} \nu^n Q_2 \ln \left(\nu\right) \left[\left(t + \frac{\mathbf{r}.\mathbf{n}}{c}\right)^2 - t_0^2\right] 
\end{eqnarray}

We construct $g_{ij}$ by substituting (\ref{eq:dphidnu})--(\ref{eq:dphidn}) into (\ref{eq:gij}). In this paper, we only require the diagonal terms of the metric. The relevant terms ($g_{00}, g_{11}, g_{22}, g_{33}$) are
\begin{eqnarray}
\nonumber g_{00} &=& \Tlag^2 \left( \frac{K_1^2}{12} +  \frac{K_1 K_2}{6} + \frac{K_2^2}{12}\right) + \Tlag^2 \left(\Tlag + 2 \frac{\mathbf{r}.\mathbf{n}}{c} + 2 \Tstart\right)  \\
\nonumber && \left(- \frac{5}{12} K_1^2 \nu^4 Q_1 - \frac{5}{12} K_1 K_2 \nu^4 Q_1 - \frac{1}{12} K_1 K_2 n \nu^{n-1} Q_2 \right.\\
\nonumber && \left. -\frac{1}{12} K_2^2 n \nu^{n-1}Q_2\right)+ \Tlag^2 \left[ 4 \Tlag^2 + 15 \left(\frac{\mathbf{r}.\mathbf{n}}{c}\right)^2 + 15 \Tlag \Tstart + 15 \Tstart^2  \right. \\
 && \left. + 15 \frac{\mathbf{r}.\mathbf{n}}{c} (\Tlag + 2 \Tstart)\right] \left(\frac{5}{36} K_1^2 \nu^8 Q_1^2 + \frac{1}{18} K_1 K_2 n \nu^{n+3} Q_1 Q_2   + \frac{1}{180} n^2 \nu^{2n-2} Q_2^2\right)
\end{eqnarray}
\begin{eqnarray}
 g_{11} &=& \frac{K_1^2 \nu^{10}}{180} \left[4 \Tlag^4 + 15 \Tlag^2 \left(\frac{\mathbf{r}.\mathbf{n}}{c}\right)^2 + 15 \Tlag^3 \Tstart + 15 \Tlag^2 T^2_{start}  + 15 \Tlag^2 \frac{\mathbf{r}.\mathbf{n}}{c} (\Tlag + 2 \Tstart)\right]
\end{eqnarray}
\begin{eqnarray}
 g_{22} &=& \frac{K_2^2 \nu^{2n}}{180} \left[4 \Tlag^4  + 15 \Tlag^2 \left(\frac{\mathbf{r}.\mathbf{n}}{c}\right)^2 + 15 \Tlag^3 \Tstart + 15 \Tlag^2 \Tstart^2  + 15 \Tlag^2 \frac{\mathbf{r}.\mathbf{n}}{c} (\Tlag + 2 \Tstart)\right]
\end{eqnarray}
\begin{eqnarray}
  g_{33} &=& \frac{K_2^2 \log \left(\nu\right)^2 \nu^{2n} Q_2^2}{180} \left[4 \Tlag^4  + 15 \Tlag^2 \left(\frac{\mathbf{r}.\mathbf{n}}{c}\right)^2 + 15 \Tlag^3 \Tstart   + 15 \Tlag^2 T^2_{start} +  15 \Tlag^2 \frac{\mathbf{r}.\mathbf{n}}{c} (\Tlag + 2 \Tstart)\right].
\end{eqnarray}

\section*{Appendix B: Semi-coherent metric}
In this appendix, we list in full the diagonal terms of the semi-coherent metric presented in (\ref{eqn:s00})--(\ref{eqn:s33}).
The relevant terms ($s_{00}, s_{11}, s_{22}, s_{33}$) are
\begin{eqnarray}
\nonumber s_{00} &=& \Tlag^2 \left( \frac{1}{12} K_1^2 + \frac{1}{6} K_1 K_2 + \frac{1}{12} K_2^2\right) \\
\nonumber && + \Tlag^2 \left(\Tlag + 2 \frac{\mathbf{r}.\mathbf{n}}{c} + \Tobs\right) \left( -\frac{5}{12} \nu^4 Q_1 K_1^2 - \frac{5}{12} K_1 K_2 \nu^4 Q_1  - \frac{1}{12} K_1 K_2 n \nu^{n-1}Q_2 - \frac{1}{12} K_2^2 n \nu^{n-1}Q_2\right)  \\
\nonumber && + \Tlag^2 \left[ 8\Tlag^2 + 30 \left( \Tlag \frac{\mathbf{r}.\mathbf{n}}{c} + \frac{\mathbf{r}.\mathbf{n}}{c}^2 + \frac{\mathbf{r}.\mathbf{n}}{c} \Tobs\right) +  15 \Tlag \Tobs + 10 \Tobs^2\right]\\
&& \left( \frac{5}{72} K_1^2 \nu^8 Q_1^2 - \frac{1}{36} K_1 K_2 n \nu^{n+3} Q_1 Q_2 + \frac{1}{360} K_2^2 n^2 \nu^{2n-2} Q_2^2\right)\\
 s_{11} &=& \frac{K_1^2 \nu^{10}}{360} \left[8 \Tlag^4 + 30 \Tlag^2 \frac{\mathbf{r}.\mathbf{n}}{c}^2 + 15 \Tlag^3 \left(2 \frac{\mathbf{r}.\mathbf{n}}{c} + \Tobs\right) + 30 \Tlag^2 \Tobs \frac{\mathbf{r}.\mathbf{n}}{c}  + 10 \Tlag^2 T^2_\text{obs} \right]\\
 s_{22} &=& \frac{K_2^2 \nu^{2n}}{360} \left[8 \Tlag^4 + 30 \Tlag^2 \frac{\mathbf{r}.\mathbf{n}}{c}^2 + 15 \Tlag^3 \left(2 \frac{\mathbf{r}.\mathbf{n}}{c} + \Tobs\right) + 30 \Tlag^2 \Tobs \frac{\mathbf{r}.\mathbf{n}}{c}  + 10 \Tlag^2 T^2_\text{obs} \right]\\
 s_{33} &=& \frac{K_2^2 \log \left(\nu\right)^2 \nu^{2n} Q_2^2}{360} \left[8 \Tlag^4 + 30 \Tlag^2 \left(\frac{\mathbf{r}.\mathbf{n}}{c}\right)^2 + 15 \Tlag^3 \left(2 \frac{\mathbf{r}.\mathbf{n}}{c} + \Tobs\right)  + 30 \Tlag^2 \Tobs \frac{\mathbf{r}.\mathbf{n}}{c}  + 10 \Tlag^2 T^2_\text{obs}\right]
\end{eqnarray}

\section*{Appendix C: Analytic accuracy estimates for the astrophysical phase model}
In this appendix, we motivate (\ref{eqn:s00})--(\ref{eqn:s33}) physically by calculating the phase error in two special cases: (i) pure gravitational-wave spin down, and (ii) pure electromagnetic spin down.
In the gravitational wave case, (\ref{eq:numodel}) reduces to

\begin{eqnarray}
\frac{d\nu}{dt} &=& -Q_1 \nu^5\\
\nu(t) &=& \frac{\nu}{(1 + 4 Q_1 \nu^4 t)^{1/4}}
\end{eqnarray}
where we take $\nu_\text{ref} = 1$\,Hz for simplicity and $\nu = \nu(t = 0)$. The gravitational wave phase at $t = \Tlag$ is then
\begin{eqnarray}
\Phi(\Tlag) - \Phi(t_0) &=& \int^{\Tlag + t_0}_{t_0} dt\, \nu(t)\\
&=&\frac{(1 + 4 Q_1 \nu^4 \Tlag)^{3/4} - 1}{3 Q_1 \nu^3}.
\end{eqnarray}

There are two regimes to be considered: (i) $\Tlag \gg (4 Q_1 \nu^4)^{-1}$, and ii) $\Tlag \ll (4 Q_1 \nu^4)^{-1}$.
In terms of the characteristic age 
\begin{equation}
\tau_c(t) = \frac{\nu(t)}{4 \lvert\dot{\nu(t)}\rvert},
\end{equation}
the two regimes correspond to (i) $\Tlag \gg \tau_c(t_0)$, and (ii) $\Tlag \ll \tau_c(t_0)$. In the case of SNR 1987A, we have $\tau_c \approx 19$\,years (in 2006, when the S5 run began) and $\Tlag \approx 1$\,hr, i.e. regime (ii).

Given small errors $\Delta Q_1$ and $\Delta \nu$ in $Q_1$ and $\nu$, the phase error that accumulates between the template and the signal after a time $\Tlag$ is
\begin{eqnarray}
\Delta \Phi &=& \frac{d\Phi}{dQ_1} \Delta Q_1 + \frac{d\Phi}{d\nu} \Delta \nu\\
&=& -\frac{1}{2}\nu^5 \Tlag^2 \Delta Q_1 + \Tlag \Delta \nu
\end{eqnarray}

Overall, therefore, the number of templates required scales as $\Tlag^3$ regardless of how rapidly the neutron star is spinning down. This scaling matches the conventional Taylor expansion if $\nu$ and $\dot{\nu}$ suffice to track the signal ($N_\text{total} \propto \Tlag^3$) but is much more economical if $\ddot{\nu}$ is needed ($N_\text{total} \propto \Tlag^6$), which happens for $\nu > 1.7 \times 10^{-5}\,\text{Hz} \left( \tau_c/10^2\,\rm{yr}\right)^{-2} \left( \Tlag/1\,\rm{hour}\right)^{-3}$. 
 In the SNR 1987A search, we cover frequencies above 0.1\,kHz, so $\ddot{\nu}$ always contributes significantly. Hence the phase model (\ref{eq:numodel}) is always preferable.

Now suppose the electromagnetic term dominates. Equation (\ref{eq:numodel}) with $\nu_\text{ref} = 1$\,Hz reduces to

\begin{eqnarray}
\frac{d\nu}{dt} &=& -Q_2 \nu^n\\
\nu(t) &=& \frac{\nu}{\left[1 + \left(n-1\right) Q_2 \nu^{n-1} t\right]^{1/(n-1)}}
\end{eqnarray}
and the gravitational wave phase after a time $\Tlag$ is
\begin{eqnarray}
\Phi(\Tlag) &=& \int^{\Tlag + t_0}_{t_0} dt\, \nu(t)\\
&=&\frac{\left[1 + \left(n-1\right) Q_2 \nu^{n-1} \Tlag\right]^{\frac{2-n}{1-n}} - 1}{\left(n-2\right) Q_2 \nu^{n-2}}.
\end{eqnarray}

For small errors $\Delta Q_2$, $\Delta \nu$ and $\Delta n$ in $Q_2$, $\nu$ and $n$, the phase error between the template at the signal after a time $\Tlag$ is
\begin{eqnarray}
\Delta \Phi &=& \frac{d\Phi}{dQ_2} \Delta Q_2 + \frac{d\Phi}{d\nu} \Delta \nu + \frac{d\Phi}{dn} \Delta n\\
\nonumber &=& -\left(\frac{\nu^n \Tlag^2}{2}\right) \Delta Q_2  + \Tlag \Delta \nu + \\
&&\Tlag \log\left[1 + \left(n-1\right) Q_2 \Tlag \nu^{n-1}\right] \Delta n
\end{eqnarray}


Hence in the electromagnetic limit, the phase error due to $\Delta \nu$ scales in the same way as in the gravitational wave limit. The phase error due to $\Delta Q_2$ scales as $\Tlag^2$, and the phase error due to $\Delta n$ scales as $\Tlag \log\left[1 + \left(n-1\right) Q_2 \Tlag \nu^{n-1}\right]$. Overall, the number of templates required scales as $\Tlag^4 \log(\Tlag)$. This represents a saving if the second frequency derivative is important which, as shown above, is true for the range of signal frequencies considered in this search.

\end{onecolumn}

\label{lastpage}
\end{document}